\documentclass[]{pasj02} 
\usepackage[switch,mathlines]{lineno} 
\usepackage{bm,url,color}

\jyear{2024}
\Received{}
\Accepted{}

\begin{document} 

\title{SIRIUS: Dark matter cusp evolution in dense dwarf galaxies}

\author{
 Katsuhiro \textsc{Kaneko},\altaffilmark{1}
 Takayuki R \textsc{Saitoh},\altaffilmark{2,3}\orcid{0000-0001-8226-4592}
 Yutaka \textsc{Hirai},\altaffilmark{4}\orcid{0000-0002-5661-033X}
 and 
 Michiko S \textsc{Fujii}\altaffilmark{1}\altemailmark\orcid{0000-0002-6465-2978} \email{fujii@astron.s.u-tokyo.ac.jp}
}
\altaffiltext{1}{Department of Astronomy, Graduate School of Science, The University of Tokyo, 7-3-1 Hongo, Bunkyo-ku, Tokyo 113-0033, Japan}
\altaffiltext{2}{Department of Planetology, Graduate School of Science, Kobe University, 1-1 Rokkodai-cho, Nada-ku, Kobe, Hyogo 657-8501, Japan}
\altaffiltext{3}{Center for Planetary Science (CPS), Graduate School of Science, Kobe University, 1-1 Rokkodai, Nada-ku, Kobe, Hyogo 657-8501, Japan}
\altaffiltext{4}{Department of Community Service and Science, Tohoku University of Community Service and Science, 3-5-1 Iimoriyama, Sakata, Yamagata 998-8580, Japan}




\maketitle

\begin{abstract}
Dwarf galaxies have a wide variety of structures, such as dark matter (DM) distribution, stellar-to-halo mass ratio, and stellar density. Recent high-resolution simulations have shown a variety of stellar-to-halo mass ratios for dwarf galaxies with a DM halo mass of $\sim 10^9 M_{\odot}$ at $z=0$. 
In this study, we performed cosmological $N$-body/smoothed-particle hydrodynamic zoom-in simulations of dwarf galaxies with the highest gas and DM particle mass resolutions of 2.37 $M_{\odot}$ and 12.8 $M_{\odot}$, respectively. The stellar-to-DM halo mass ratio of one of our simulated dwarf galaxies was $\sim 10^{-4}$, typical for satellites of the Milky Way. 
The stellar mass ($10^5 M_{\odot}$) and half-mass radius (68\,pc) were also similar to those of the satellites of the Milky Way. The power-law slope of the DM halo was $\alpha = -1.1$. 
On the other hand, the other simulated galaxy exhibited a stellar-to-halo mass ratio of $\sim 10^{-3}$ and a steeper power-law slope ($\alpha=-1.9$) than the other; the presence of baryonic matter deepened the cusp. The mass of $>10^6 M_{\odot}$ and a half-mass radius of $\sim 36$\,pc of this galaxy were similar to those of ultra-compact dwarf galaxies rather than the satellites of the Milky Way. This DM halo grew in mass earlier than the former one, and the central DM density was higher than that of the other even in the DM-only simulations. 
\end{abstract}

\section{Introduction}
The $\Lambda$ cold dark matter ($\Lambda$CDM) model has been widely accepted as the favored model to explain the formation of the large-scale structure of the Universe \citep{Tegmark_2004,Springel_2005a}. 
Stars are formed in such dark matter (DM) halos, and  galaxies evolve inside DM halos. The stellar mass depends on the halo mass, and the relation has been derived using abundance matching (e.g., \cite{2013ApJ...770...57B}).

However, at the low-mass end $(M_{\rm h}\lesssim 10^9 M_{\odot})$, which corresponds to the mass range of ultra-faint dwarfs, the stellar-to-halo mass ratio is still hardly determined from observations. The relation extrapolated from abundance matching of more massive galaxies shows a variation in the stellar-to-halo mass ratio \citep{2013ApJ...770...57B,2013MNRAS.428.3121M,2022NatAs...6..897S}. 
On the other hand, numerical studies have also been used to determine the stellar-to-halo mass ratio. Recent highest resolution simulations have reached the resolution of $4 M_{\odot}$ for gas \citep{LYRA}. However, the number of samples is not large. Lower mass resolution simulations with a few thousand $M_{\odot}$ for gas particles have more samples, but their results do not agree with each other \citep{Hopkins_2018,GEAR,2022NatAs...6..897S}. 

The structures of dwarf galaxies have also not yet been fully understood. 
The density profile of DM halos obtained from numerical simulations assuming the $\Lambda$CDM model showed a cuspy density profile in the central region ($\rho \propto r^{-1}$) \citep{Navarro_1996}. 
On the other hand, some observations of dwarf galaxies suggested a flat core ($\rho \sim $ const.) in the innermost region of the DM halo \citep{Flores_1994,Moore_1994,1995ApJ...447L..25B,2011AJ....141..193O} and known as the core-cusp problem \citep{deBlok_2010}.

Baryonic physics, such as star formation and supernova feedback, is one possible mechanism for creating a flat core. \citet{Governato_2010} showed the formation of a core profile in cosmological hydrodynamic simulations. In their simulations, repeated gas accretion and removal due to supernova feedback after bursty star formation caused oscillations of gas potential and, as a result, transformed the cuspy density profile of the halo into a core \citep{2012MNRAS.421.3464P,2014ApJ...793...46O,2017ARA&A..55..343B}. \citet{2014ApJ...793...46O} demonstrated that the repeated oscillation in the central region is crucial in causing the transition from a cusp to a core using their analytic calculations and $N$-body simulations.

Such a transition could occur in relatively massive dwarf galaxies. 
\citet{2014MNRAS.437..415D} reported that the cusp-core transition occurs when the stellar-to-halo-mass ratio is $\sim 10^{-3}$, but not when it is $\sim 10^{-4}$. Similar results were obtained in other simulation studies \citep{Tollet_2016,Lazar_2020}.
However, the halo mass was more than $\sim 10^{10} M_{\odot}$ in these studies. In smaller galaxies, the duration of star formation is short, and therefore, the feedback energy may not be sufficient to form a DM core.

For lower mass dwarf galaxies, a higher resolution is required to resolve the small scale down to the central region. 
\citet{Wheeler_2019} performed cosmological zoom-in simulations for $2$--$9\times 10^9M_{\odot}$ dwarfs. Their mass resolution reached $30 M_{\odot}$, whereas the mass resolution of $>10^{10} M_{\odot}$ galaxies in previous studies were $>10^3 M_{\odot}$ \citep{2014MNRAS.437..415D}. The gravitational softening length was 14\,pc. In their simulations, one of their dwarfs close to $10^{10}M_{\odot}$ achieved a stellar-to-halo mass ratio of $>10^{-4}$ and had a core. This dwarf galaxy continued star formation beyond the reionization.  

Another series of cosmological zoom-in simulations, EDGE, has been performed for DM halos with $\sim 10^9$--$10^{10} M_{\odot}$ at $z=0$ using an adaptive mesh refinement (AMR) hydrodynamics
code \citep{2020MNRAS.491.1656A}. The finest spatial resolution reached 3\,pc. Their mass resolution for star particles was $\sim 20M_{\odot}$.
In \citet{2021MNRAS.504.3509O_EDGE}, they reported that mergers formed a core rather than repeated feedback and gas inflow. 
They also found a transition from cusp to core depending on the stellar mass fraction formed before and after the reionization \citep{2025MNRAS.536..314M}. These results suggest that the long-term evolution of galaxies is essential for the DM halo structures of this mass range. 

The recent highest resolution simulation, LYRA \citep{LYRA}, also investigated the power of the DM distribution. Their simulation can resolve down to $\sim 1$\,pc. Such a high resolution can resolve feedback due to the supernovae without a subgrid model \citep{2019MNRAS.483.3363H}, different from low-resolution simulations \citep{GEAR,Hopkins_2018}. 
They did not find a correlation between the star formation activity and the shape of the DM halos in their simulations. They also reported that one of their models showed a central cusp deeper than that of their dark-matter only (DMO) simulation \citep{2025arXiv251110582S_LYRA}. 

On the other hand, recent observations have revealed a large scatter in the power of the central regions of the DM halos for the Milky Way (MW) satellite galaxies \citep{Hayashi_2020,2023ApJ...953..185H}. Although the uncertainty is also large, the observed shape of the DM halos does not seem to depend on the stellar-to-halo mass ratio for dwarf galaxies. 
\citet{2015MNRAS.452.3650O} also showed that a diversity in the circular velocity of dwarf galaxies in their inner region compared to their simulations. This suggests that a diversity in DM halo distribution in dwarf galaxies.
Thus, the structures of DM halos of dwarf galaxies, especially low-mass ones, still have uncertainties.

In this paper, we present the results of relatively low-mass ($\sim 10^9 M_{\odot}$ in DM) dwarf galaxy formation simulations with a mass resolution comparable to that achieved by \citet{Wheeler_2019} and a spatial resolution comparable to that achieved by \citet{2020MNRAS.491.1656A} to investigate the DM halo and stellar distributions. We have further resolved individual stars. 
The structure of this paper is as follows: Section 2 provides a description of our model and simulation setup. Section 3 presents the evolution of the cusp in our simulated dwarf galaxies. A comparison with observations and previous studies has been presented in Section 4. Section 5 summarizes this study.

\section{Methods}
\subsection{Simulation Code}
We performed the cosmological zoom-in simulations of dwarf galaxies using an $N$-body/smoothed-particle hydrodynamics (SPH) code, \textsc{ASURA+BRIDGE} \citep{Saitoh_2009,Hirai_2021,Fujii_2021}. 
In this code, gravity was calculated using the tree method \citep{Barnes_1986}, and an opening angle of $\theta=0.5$ was adopted. The hydrodynamics was computed using the density-independent SPH (DISPH) method \citep{Saitoh_2013}. 
For radiative cooling and heating of the gas,  metallicity-dependent cooling and heating functions from $10\,\mathrm{K}$ to $10^9\,\mathrm{K}$ generated by Cloudy ver. 13.05 \citep{Ferland_1998,Ferland_2013,Ferland_2017} were employed.
We also adopted the UV background radiation model by \citet{Haardt_2012}, combining the self-shielding model by \citet{Rahmati_2013}, and set the reionization due to UV background radiation to occur at $z=8.5$, $t=0.588\,\mathrm{Gyr}$.
The chemical evolution of gas and stars was treated using the chemical evolution library \textsc{CELib} \citep{Saitoh_2017, Hirai_2021}. 

For the gravitational softening lengths of the DM particles ($\epsilon_{\mathrm{DM}}$), we followed \citet{Hopkins_2018}:
\begin{equation}
    \epsilon_{\mathrm{DM}} = 30\,\mathrm{pc} \, \left(\frac{m_{\rm DM}}{1000\,M_{\odot}}\right)^{\frac{1}{2}} \left(\frac{M_\mathrm{vir}}{10^{12}\,M_{\odot}}\right)^{-0.2},
\label{equation:SoftningDM}
\end{equation}
where $m_{\rm DM}$ is the mass of the DM particles and $M_\mathrm{vir}$ is the virial mass of the system. A constant value of $M_\mathrm{vir}=10^9\,M_{\odot}$ was used in our simulations.
For those of gas, we followed \citet{Dutton_2020} but modified the factor:
\begin{equation}
    \epsilon_{\mathrm{Gas}} = 2.13\,\mathrm{pc} \left(\frac{m_\mathrm{gas}}{1 M_\odot }\right)^{\frac{1}{3}} \left(\frac{\rho_\mathrm{th}}{100\,\mathrm{cm}^{-3}}\right)^{-\frac{1}{3}}, \label{equation:SoftningGas}
\end{equation}
where $m_\mathrm{gas}$ is the initial gas mass and $\rho_\mathrm{th}$ is the threshold density of star formation (see below). Initially, all the gas particles had the same mass, and the stars born in the simulations took over the softening length of the gas. The softening length for stars was set to be equal to that for the gas particles ($\epsilon_{\rm Star}=\epsilon_{\rm Gas}$). 
In the gravity calculation, the gravitational softening lengths ($\epsilon_{\mathrm{soft}}^{\mathrm{calc}}$) of DM and gas/star were varied by the redshift $z$,
\begin{equation}
\epsilon_{\mathrm{soft}}^{\mathrm{calc}} =
\begin{cases}
\epsilon_{\mathrm{soft}},\,(z < z_\mathrm{f}), &\\
\epsilon_{\mathrm{soft}} \times \frac{1+z_\mathrm{f}}{1+z}\,(z \geq z_\mathrm{f}), &
\end{cases}
\label{equation:SoftningRedshift}
\end{equation}
where $\epsilon_{\mathrm{soft}} = \epsilon_{\mathrm{soft}}^{\mathrm{DM}}$ or $ \epsilon_{\mathrm{soft}}^{\mathrm{Gas,Star}}$ and we employed $z_\mathrm{f}=9.0$.

\subsection{Star formation and initial mass function}
We adopted a star-by-star star-formation method \citep{Hirai_2021,Hirai_2025}, in which individual star particles have different masses following a given initial mass function (IMF). 
The formation of a star occurred when a gas particle satisfied the following conditions: (1) the divergence of velocity was less than 0 ($\nabla \cdot \bm{v}<0$), (2) the gas number density was greater than $100\,\mathrm{cm}^{-3}$, and (3) the temperature was less than $1000\,\mathrm{K}$.

The star formation was assumed to probabilistically occur following the Schmidt law \citep{Schmidt_1959,Katz_1992}:
\begin{equation}
    \frac{d\rho_{\star}}{dt} = - \frac{d\rho_{\mathrm{gas}}}{dt} = c_{\star} \frac{\rho_{\mathrm{gas}}}{t_{\mathrm{dyn}}},
\label{equation:Schmidt}
\end{equation}
where $\rho_\star$, $\rho_{\mathrm{gas}}$ are a density of star and gas, respectively, $c_{\star}$ is a dimensionless star-formation efficiency, and $t_{\mathrm{dyn}}=\left(4\pi G \rho_{\mathrm{gas}}\right)^{1/2}$ is a local dynamical time of the star formation region. 
From this relation, the probability of a gas particle forming a star in a time range of $\Delta t$ was calculated as 
\begin{equation}
    p=\frac{m_{\mathrm{gas}}}{\langle m_{\star} \rangle} \left[1-\exp\left(-c_{\star} \frac{\Delta t}{t_{\mathrm{dyn}}}\right)\right],
\label{equation:SFProbability}
\end{equation}
where $\langle m_{\star} \rangle$ is the average stellar mass of the IMF.
A random number $0\leq R < 1$ was generated when a gas particle satisfied the three criteria. If $R<p$,  a stellar particle was created (see \cite{Hirai_2021} for more details). 
In this study, we adopted $c_{\star}=0.5$ following previous simulations by \citet{Saitoh_2008}.

Once a star-forming region was identified, a stellar mass was drawn following a given mass function. We adopted a Salpeter IMF \citep{Salpeter_1955} but modified it to reduce the number of particles of low-mass stars. We adopted a modified IMF (hereafter cutoff IMF), in which stars with $m_{\star}<1M_{\odot}$ were summarized as superparticles with $1.0\leq m_{\star}/M_{\odot} \leq 1.1$. The mass function was expressed as:
\begin{equation}
\frac{dN}{d\log_{10} m_{\star}} \propto
\begin{cases}
m_{\star}^{-1.35}, & \\
\hspace{2.5cm}(1.1 < m_{\star}/M_{\odot} <120), &\\
\frac{1}{1.1 - 1.0\,M_{\odot}} \int_{0.1\,M_{\odot}}^{1.1\,M_{\odot}} m_{\star}^{-1.35} dm_{\star}, &\\
\hspace{2.5cm}(1.0 \leq m_{\star}/M_{\odot} \leq 1.1). &
\end{cases}
\label{equation:CutoffIMF}
\end{equation}
With this cutoff IMF, the number of low-mass stars was reduced, thus maintaining the mass fraction of massive stars in the entire mass distribution.

The stellar mass was also limited by the local gas mass to conserve the local mass. If the mass drawn from the IMF was smaller than that of the gas particle, the star could be born from the gas particle. Even if the stellar mass was larger than the mass of the parental gas particle, we searched the gas mass within $r_{\mathrm{max}} = 3\,\mathrm{pc}$. If the gas within $r_{\mathrm{max}}$ was twice as massive as the selected stellar mass, a star particle was formed by collecting the stellar mass from the gas particles within $r_{\mathrm{max}}$. If not, a new stellar mass was drawn. 
The center of mass of the gas particles determined the position and velocity of the star particle.  

\subsection{Stellar evolution and feedback}
Stellar evolution and feedback were also treated with \textsc{CELib}. 
Lifetimes of star particles were determined from the metallicity-dependent stellar lifetime table in \citet{Portinari_1998}.
Instead of calculating radiation transfer, we assumed that massive stars ($>15\,M_{\odot}$) formed spherical H\textsc{ii} regions. We calculated each Str\"{o}mgren radius, which is the radius of the H\textsc{ii} region of the star, using the local gas density, and provided thermal energy to maintain the temperature within the Str\"{o}mgren radius 10,000\,K \citep{Fujii_2021}. 
The maximum extent of H\textsc{ii} regions was set to $30\,\mathrm{pc}$ from the center to prevent the H\textsc{ii} region from expanding too large in almost gas-free regions in the simulation \citep{Fujii_2021}.

Following the yield table of \citet{Nomoto_2013}, stars within the mass range of $13\leq m_{\star}/M_{\odot} \leq 40$ exploded as core-collapse supernovae when they reached the end of their lives.
We assumed that these supernovae released energy of $10^{51}\,\mathrm{erg}$.
We also assumed that $5\,\%$ of stars of $20-40\,M_{\odot}$ exploded as broad-line Type Ic supernovae (hypernovae) and released energy of $10^{52}\,\mathrm{erg}$.
We adopted a simple supernova feedback model that injects all supernova energy as thermal energy into the surrounding gas particles. 
We did not include other stellar feedback such as Type Ia supernovae because the gas was removed mainly due to reionization before such feedback is activated.

\subsection{Initial Conditions}

As a first step, we performed a DM-only cosmological simulation of a $4\,\mathrm{Mpc}\,h^{-1}$ box. The initial condition was generated by \textsc{MUSIC} \citep{Hahn_2011}.
We assumed a flat $\Lambda$CDM cosmology and set cosmological parameters as the values obtained from \citet{Planck_2020} ($H_0=67.32$, $\Omega_{\mathrm{m} 0}=0.3158$, $\Omega_{\Lambda 0}=0.6842$, and $\Omega_{\mathrm{b} 0}=0.04939$). The mass resolution of the low-resolution cosmological simulation was $m_{\mathrm{DM}} = 5\times10^{5}\,M_{\odot}$.
Starting from $z=100$, we performed a cosmological simulation using \textsc{GADGET-2} \citep{Springel_2005b} up to $z=0$.
Amiga's Halo Finder (AHF; \cite{Knollmann_2009}) was used to detect DM halos at $z=0$, and then the DM halos with $M_{\mathrm{halo}} \simeq 10^9 \, M_{\odot}$ at $z=0$ in isolated environments were selected. For the isolated condition, we adopted no more than 10 times massive DM halos within 1\,Mpc.
 
\begin{figure}
    \begin{center}
        \includegraphics[width=\columnwidth]{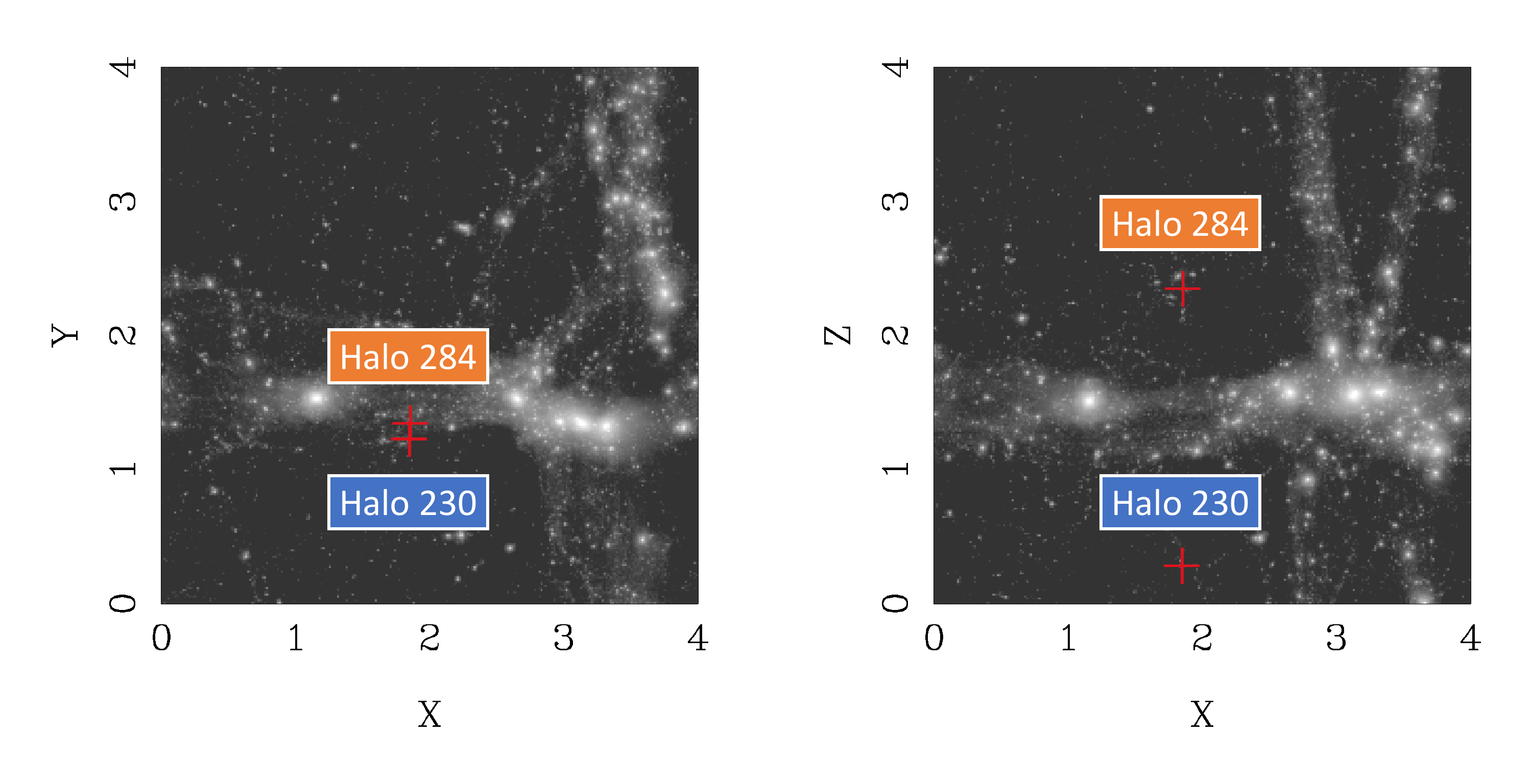}
        \includegraphics[width=\columnwidth]{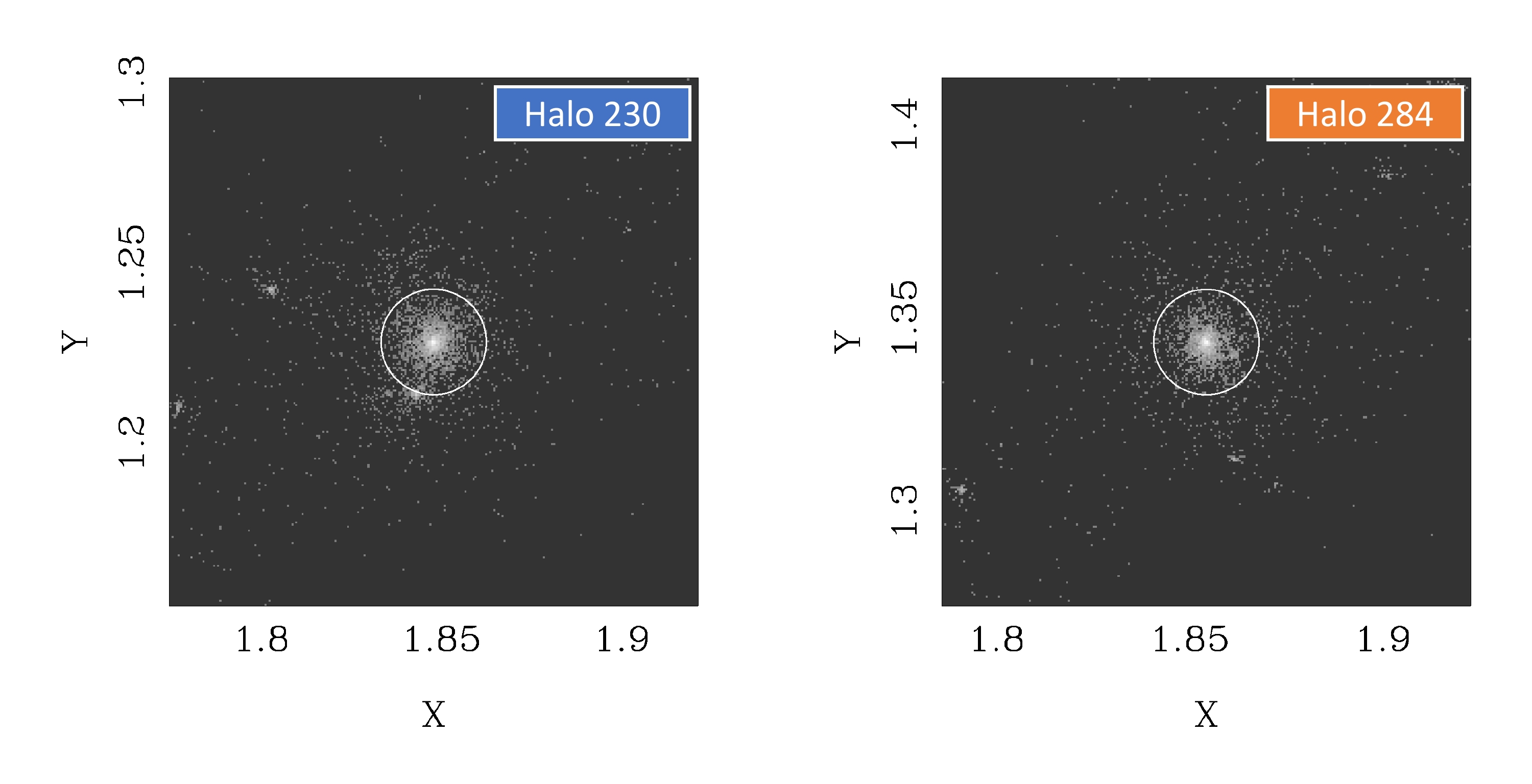}
    \end{center}
	\caption{\textit{Top panels}: Dark matter surface density distribution, and the positions of Halo 230 and Halo 284 in it. The red crosses show the centers of the halos at $z=0$. \textit{Bottom panels}: Zoomed-in images of each halo. The white circles show the virial radii.
    Axis scales are in $\mathrm{Mpc} / h$ for all panels. 
    Alt text: Four-by-four panel figure. The top panels show the positions of Halo 230 and 284 in a four comoving megaparsec square scale large-scale distribution of dark matter surface density in x-y (left) and x-z (right) planes. Bottom panels show zoomed-in images of Halo 230 (left) and Halo 284 (right).
    }
	\label{fig:GadgetCosmo}
\end{figure}

Using \textsc{MUSIC}, the refined initial conditions of the selected halos were generated.
The zoom-in region was determined by the following method (cf. \cite{Griffen_2016}): (1) Pick up particles within $4\,R_{\mathrm{vir}}$ from a halo center at $z=0$. (2) Determine an ellipsoid region such that all of these particles at $z=100$ are included. (3) The region whose size is multiplied by $1.05$ is the zoom-in region for reducing the contamination of the coarse boundary particles.
Low-resolution zoom-in simulations were performed with $m_{\mathrm{gas}}\sim1.2\times10^{3}\,M_{\odot}$ and $m_{\mathrm{DM}}\sim6.5\times10^{3}\,M_{\odot}$ using \textsc{ASURA+BRIDGE}. Hereafter, we refer to them as Hydro Low models. From the results of this test, the most and least massive dwarf galaxies in stellar mass (Halo 284 and Halo 230, respectively) were chosen for higher-resolution simulations. Their surroundings and the DM distributions are shown in Figure~\ref{fig:GadgetCosmo}. 
Table~\ref{table:HaloPropertyGadget} summarizes the properties of Halo 230 and Halo 284 at $z=0$ obtained with AHF.

\begin{table}
 \caption{Halo properties at $z=0$. Note that $V_{\mathrm{max}}$ is the maximum circular velocity.}
 \label{table:HaloPropertyGadget}
 \centering
  \begin{tabular}{cccc}
   \hline
    Halo Number & $M_{\mathrm{vir}}\,[M_{\odot}]$ & $R_{\mathrm{vir}}\,[\mathrm{kpc}]$ &  $V_{\mathrm{max}}\,[\mathrm{km}/\mathrm{s}]$ \\
   \hline \hline
   230 & $1.075 \times 10^9$ & $21.75$ & $18.54$ \\
   284 & $8.78 \times 10^8$ & $20.29$ & $17.70$ \\
   \hline
  \end{tabular}
\end{table}

For Halo 230 and Halo 284, high-resolution zoom-in simulations were performed with $m_{\mathrm{gas}}=19.0\,M_{\odot}$ and $ m_{\mathrm{DM}}=102\,M_{\odot}$, and we refer to them as Hydro models. The Hydro model simulation was continued up to $z=0.5$ for Halo 230, whereas for Halo 284, it was continued until $z=0.32$ because it experienced a merger around $z=0.5$. For comparison, we also performed simulations without gas, and we refer to them as DM-only (DMO) models. The mass resolution of this simulation was the same as that of the Hydro models. The resolution is summarized in Table~\ref{table:resolution}.

In addition, we performed a higher-resolution simulation for only Halo 284 up to $z=8.72$, which we refer to as the Hydro High model (see Appendix~\ref{sec:resolution}). The mass resolution of this model was $2.37M_{\odot}$, which was sufficiently high to resolve the supernova explosions \citep{2019MNRAS.483.3363H}. Due to our limited computational resources, we terminated this simulation at $z=8.72$ and investigated the convergence at this point in time.

\begin{table}
 \caption{Resolution of the simulations}
 \label{table:resolution}
 \centering
  \begin{tabular}{lcccc}
   \hline
    Model & $m_{\mathrm{gas}}$ & $\epsilon_{\mathrm{gas}}$ &  $m_{\mathrm{DM, min}}$ & $\epsilon_{\mathrm{DM, min}}$\\
     & [$M_{\odot}$] & $[\mathrm{pc}]$ &  $[M_{\odot}]$ & $[\mathrm{pc}]$\\
   \hline \hline
   DMO & - & - & $1.21\times10^2$ & 13.8 \\
   Hydro Low & $1.21\times10^3$ & 22.7 & $6.55\times10^4$ & 102 \\
   Hydro & $1.90\times10^1$ & 5.68 & $1.02\times10^2$ & 12.7 \\
   \hline
  \end{tabular}
  From left to right: (1) gas-mass resolution ($m_{\rm gas}$), (2) gas softening length ($\epsilon_{\rm gas}$) (3) minimum dark-matter mass ($m_{\rm DM, min}$) (4) minimum dark-matter softening length ($\epsilon_{\rm DM, min}$). 
\end{table}

\subsection{Determination of the halo center}
\label{section:HaloCenterSelection}
In order to determine the DM density profiles of the simulated dwarf galaxies, the shrinking sphere method \citep{Klypin_1997,Power_2003,Navarro_2004,Fitts_2017} was used.
First, a sphere with the center and virial radius calculated with AHF was considered. Next, the center-of-mass position of the DM particles within the sphere was calculated, and the center of the sphere was moved there. Then, the radius of the sphere was reduced by $2.5\%$.
This operation was repeated until the number of DM particles in the sphere was less than 1000. 
The viral mass, radius, and radial profiles were calculated using the center determined with this method.
Here, the virial radius was the radius within which all stars were bound to the halo, and the mass was the total mass within the virial radius. 

\section{Results}

\subsection{Halo mass evolution and star formation}
\begin{figure*}
    \begin{center}
    \includegraphics[width=0.9\columnwidth]{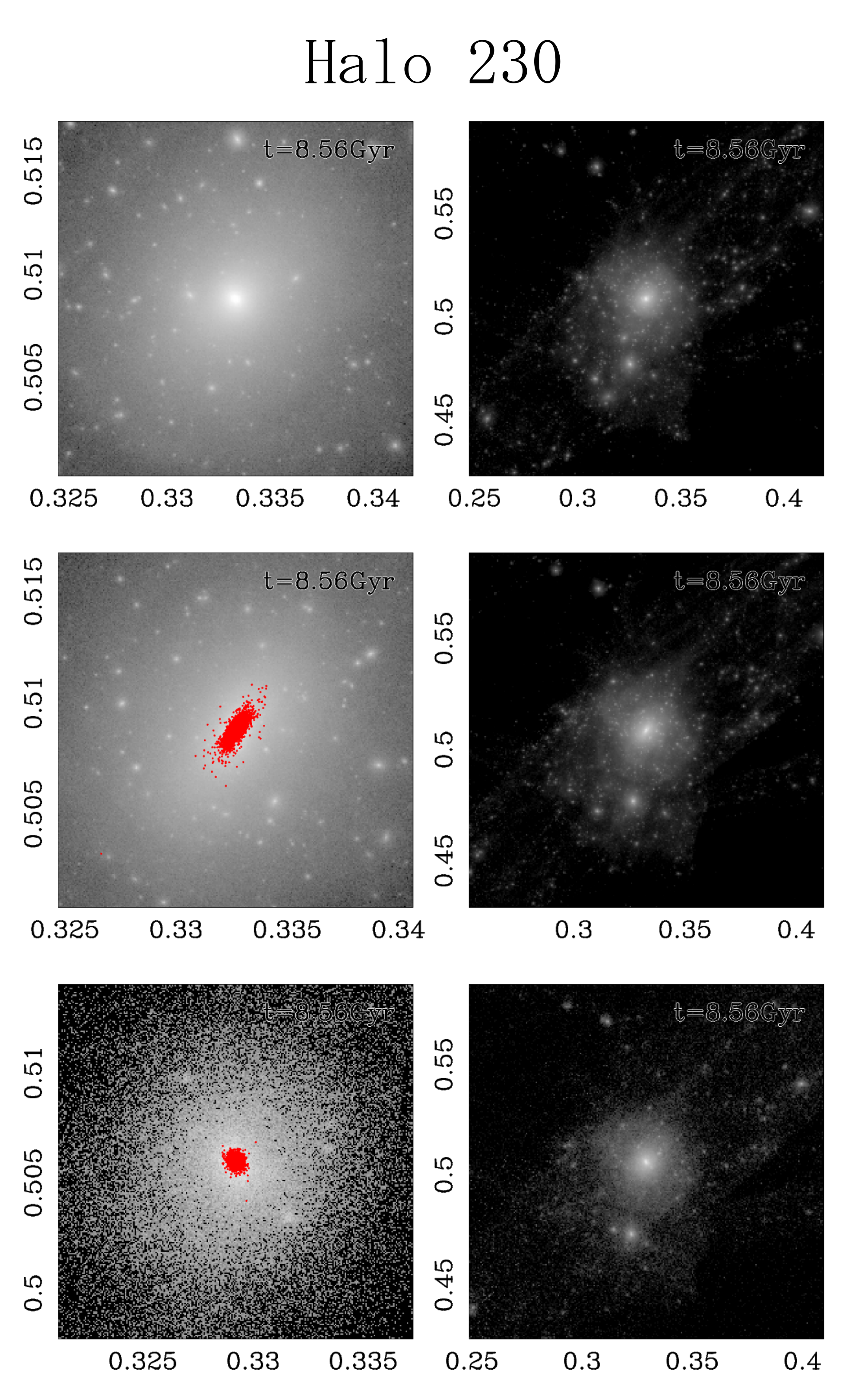}
    \includegraphics[width=0.9\columnwidth]{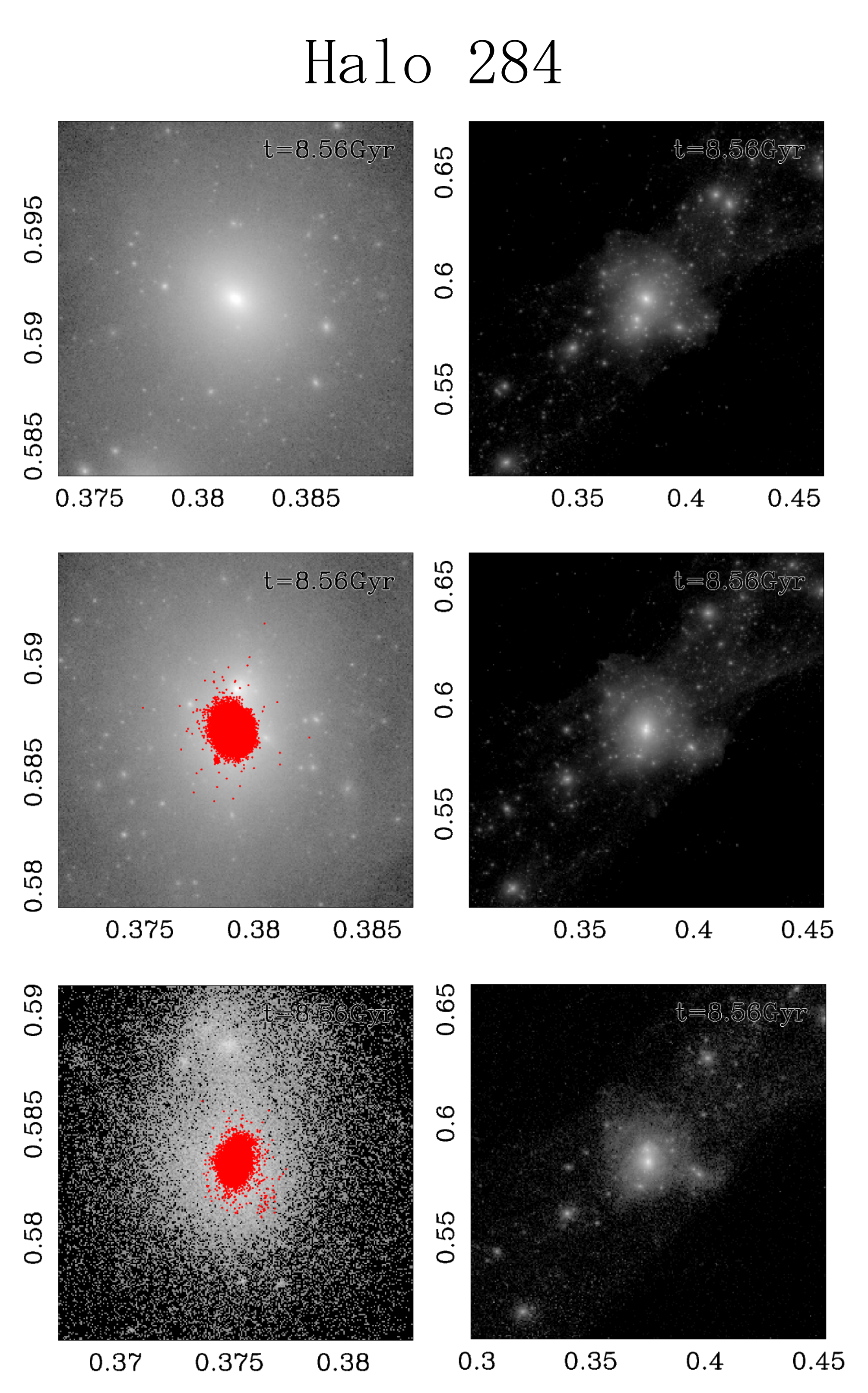}
    \end{center}
	\caption{DM surface density distributions of Halo 230 (\textit{left}) and Halo 284 (\textit{right}) at $z=0.5$. From top to bottom, each row represents the results of the DMO, Hydro, and Hydro Low models, respectively. The left and right columns of each halo show the DM surface density within $0.5\,R_{\mathrm{vir}}(t)$ and $5\,R_{\mathrm{vir}}(t)$ from the halo center, respectively. 
    The red dots in the left column indicate star particles. Axis scales are in $\mathrm{Mpc} / h$ for all panels. Alt text: Twelve-panel grayscale images of the dark matter surface density centered on the halo center. Red dots are plotted on the grayscale images for the Hydro and Hydro Low models.}
	\label{fig:SurfaceDensity}
\end{figure*}

In Figure~\ref{fig:SurfaceDensity}, we present the distribution of the DM and stars at $z=0.5$. At this time, the star formation has already finished, and the dwarf galaxies are gas-free. Stars are concentrated in the central region of the main DM halo. 

\begin{figure*}
	\begin{center}
		\includegraphics[width=0.95\columnwidth]{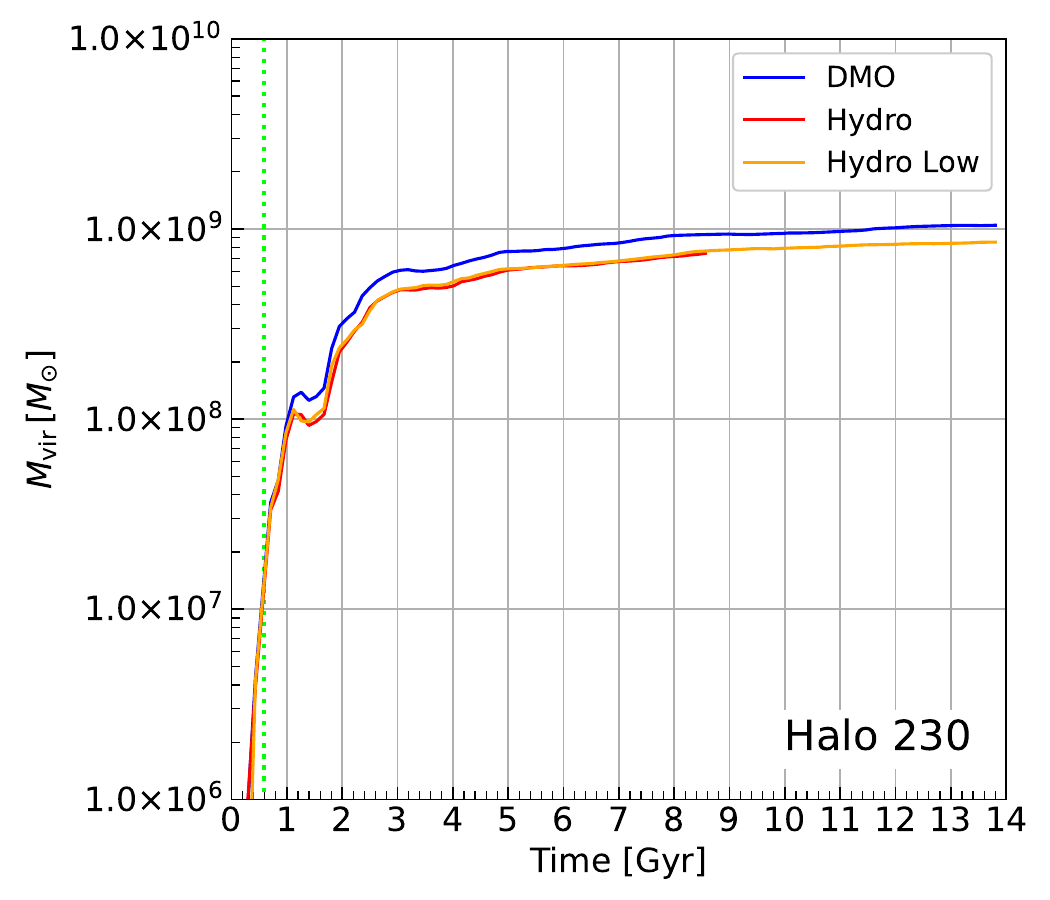}
		\includegraphics[width=0.95\columnwidth]{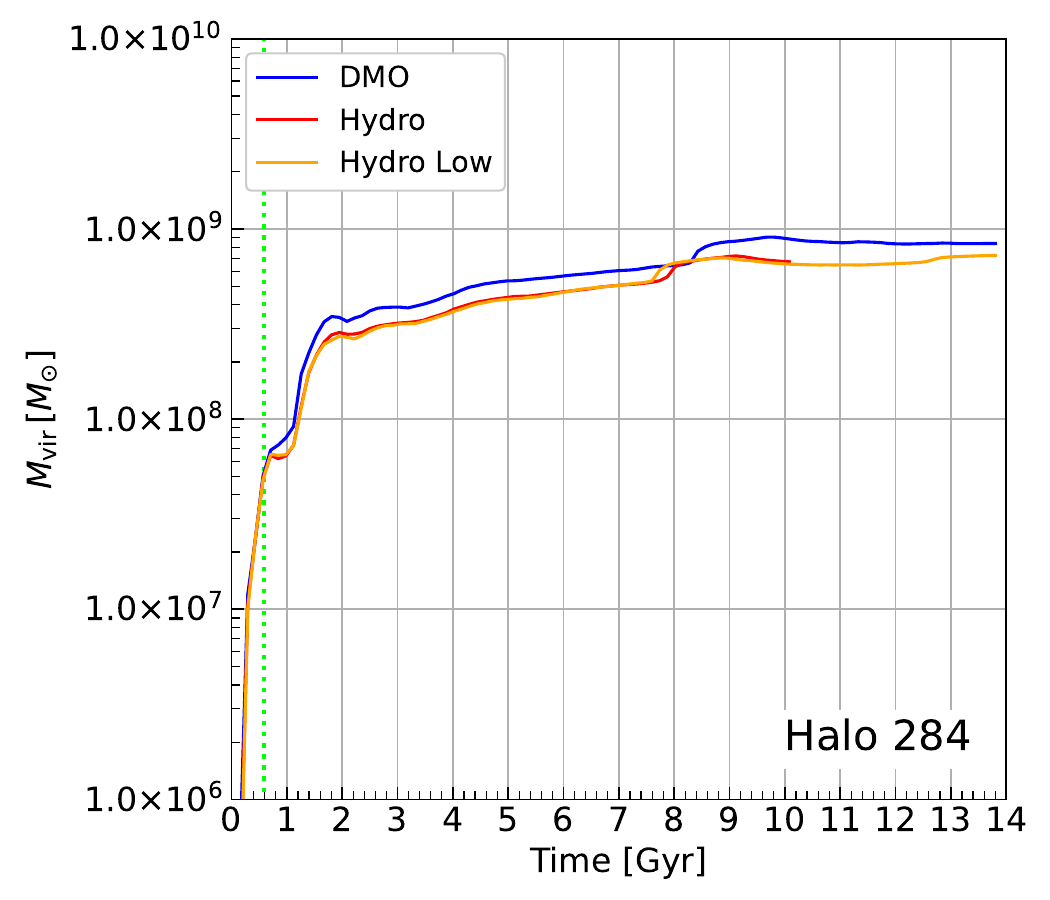}
	\end{center}
	\caption{Time evolution of the viral mass of the halo ($M_{\mathrm{vir}}$) for Halo 230 (\textit{left panel}) and Halo 284 (\textit{tight panel}). The green dotted vertical line indicates the reionization epoch assumed in this simulation ($z=8.5\,$, $t=0.588\,\mathrm{Gyr}$). Alt text: Two line graphs. The range of the x-axis (time) is zero to fourteen gigayears. Blue and red lines are for DMO and Hydro runs, respectively. The virial masses increase with time and reach the final mass at around eight point five gigayears.}
	\label{fig:AnalysisData_Time-Mvir}
\end{figure*}

The halo virial mass evolution is shown in Figure~\ref{fig:AnalysisData_Time-Mvir}. The virial mass and radius were calculated using the center derived by the method described in Section \ref{section:HaloCenterSelection}. Because of the existence of baryons, the virial masses of the Hydro and Hydro Low models are less than those of the DMO models. 
As shown in this figure, the halo virial mass and radii did not strongly depend on the resolution. 

\begin{figure*}
	\begin{center}
		\includegraphics[width=0.95\columnwidth]{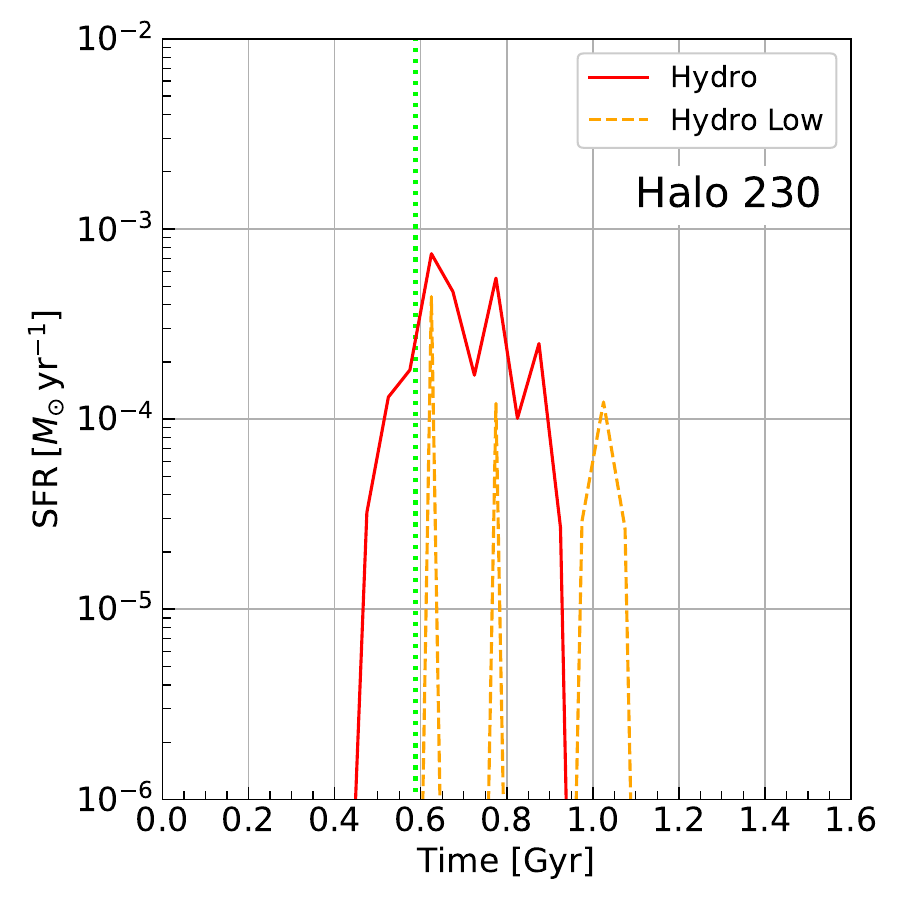}
        \includegraphics[width=0.95\columnwidth]{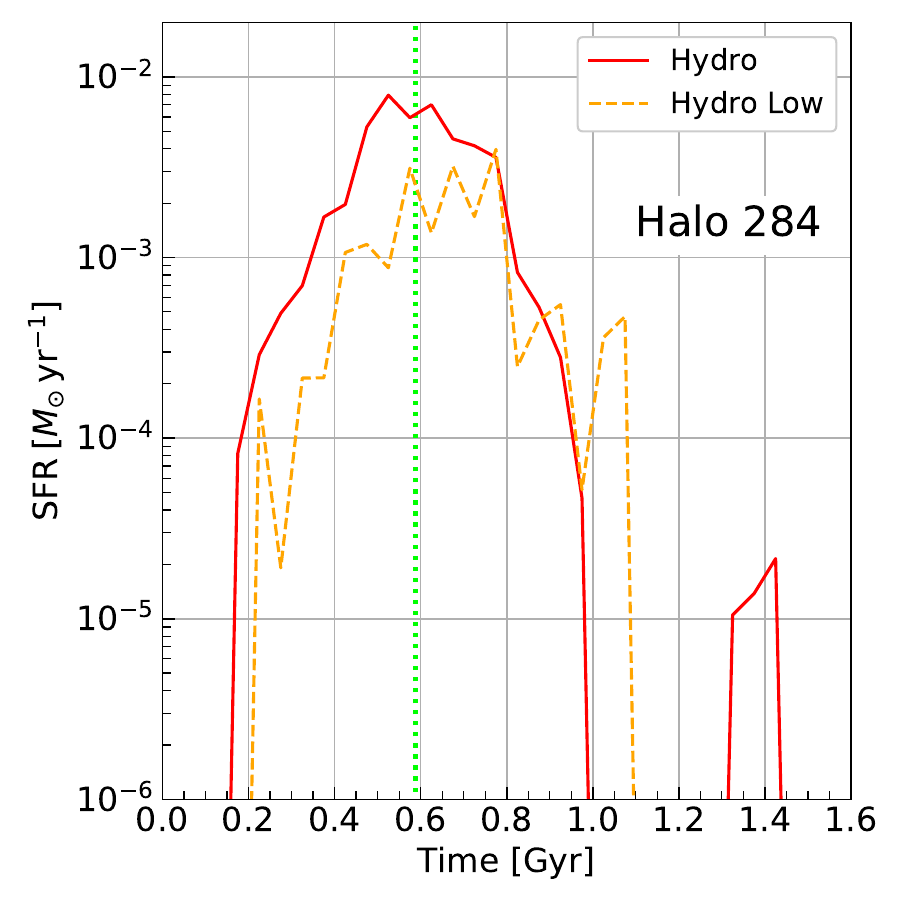}
	\end{center}
	\caption{Star formation histories of Halo 230 (left) and Halo 284 (right). Star formation rates are averaged over $50\,\mathrm{Myr}$. The green dotted line indicates the reionization epoch. Alt text: Line graphs peaked at around the reionization epoch. The red and orange lines correspond to the Hydro and Hydro Low models, respectively. The peak value of the star formation rate of Halo 230 is slightly less than ten to the minus three solar mass per year. The peak value of Halo 284 is slightly less than ten to the minus two solar masses per year. }
	\label{fig:AnalysisData_Time-SFR}
\end{figure*}

Figure~\ref{fig:AnalysisData_Time-SFR} shows the time evolution of the star formation rate averaged over $50\,\mathrm{Myr}$.
Star formation rates were calculated using an archaeological method (e.g., \cite{Fitts_2017}) as follows: First, the member stars within $0.5\,R_{\mathrm{vir}}$ of each halo at $8.56\,\mathrm{Gyr}$ for Halo 230 and $10.1\,\mathrm{Gyr}$ for Halo 284 were selected. Then, their formation epoch was calculated from their stellar age. Finally, the star formation rate of this dwarf was calculated based on the ages of its member stars. 

The Halo 230 Hydro model showed a single-peak star formation history. The star formation started earlier than the reionization epoch and ended around 1\,Gyr. No further star formation occurred beyond this time, although the halo mass continued to increase via mergers and accretions at later times (see Figure~\ref{fig:AnalysisData_Time-Mvir}). On the other hand, in the Hydro Low model, the star formation started around the reionization epoch and continued slightly later in time. The final mass and the half-mass radius of the stars were $1.23\times10^5 M_{\odot}$ and 68.3\,pc, respectively (see Table~\ref{table:HaloProperty}).

For Halo 284, the Hydro and Hydro Low models showed a single-peaked star formation history similar to that of Halo 230, but a small star formation occurred again at $\sim1.4\,\mathrm{Gyr}$. Further, around 8--10 Gyr, Halo 284 experienced a major merger, but no star formation occurred at this time. 
The star formation rate of Halo 284 was an order of magnitude higher than that of Halo 230. The final stellar mass of Halo 284 ($2.12\times 10^6M_{\odot}$) was an order of magnitude higher than that of Halo 230. However, the half-mass radius of the star for Halo 284 was only 36.4\,pc, which is smaller than that for Halo 230. Thus, Halo 284 is more compact compared to Halo 230. 

\begin{figure}
	\begin{center}
		\includegraphics[width=0.95\columnwidth]{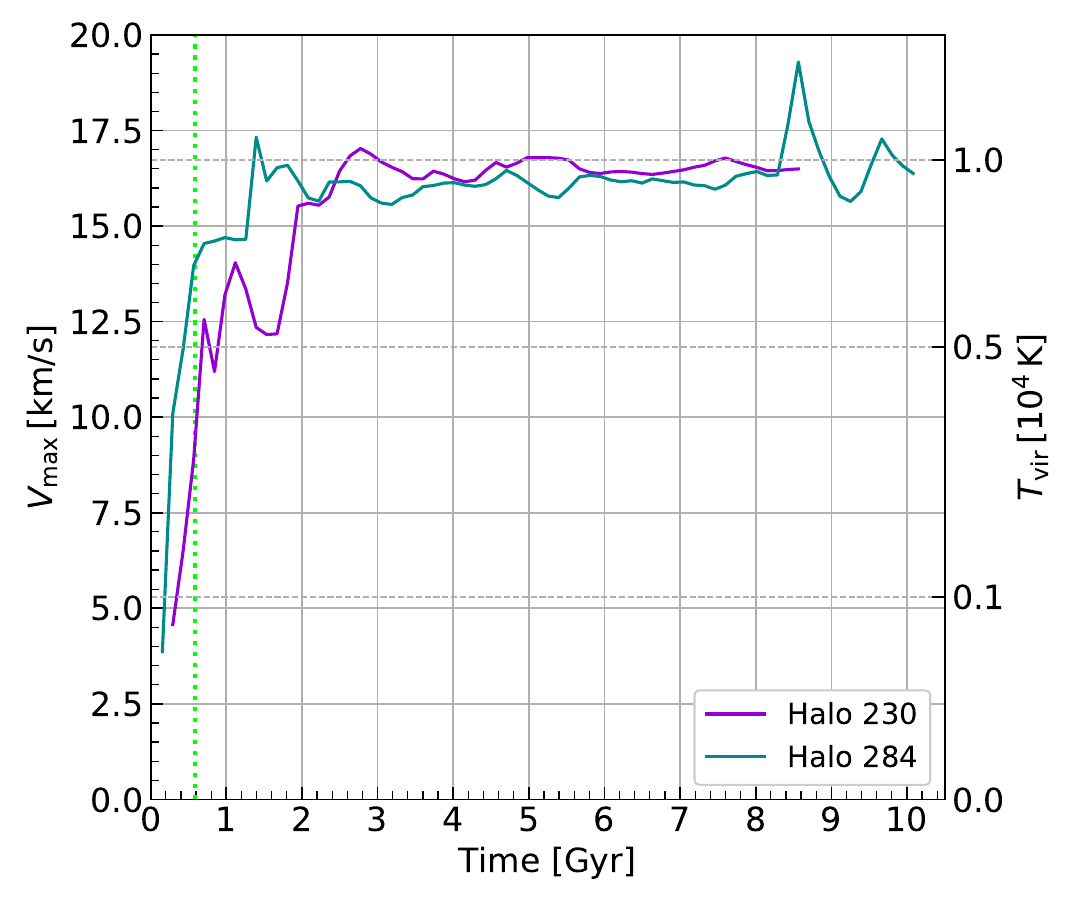}
	\end{center}
	\caption{Time evolution of $V_{\mathrm{max}}$ and $T_{\mathrm{vir}}$ in each simulation. The left axis shows the maximum circular velocity, whereas the $y$-axis tics on the right side show the virial temperature. The green dotted line represents the time at which reionization occurs in this simulation. Alt text: Line graph with two lines. Purple and green lines are for Halo 230 and 284, respectively. The X-axis indicated the time in gigayears. Left and right y-axes show the circular velocity and the corresponding virial temperature, respectively. The circular velocity increases with time and reaches a circular velocity of approximately 17 kilometers per second. The circular velocity corresponds to ten thousand kelvin. }
	\label{fig:Vmax_Tvir}
\end{figure}

The difference between Halo 230 and Halo 284 appears to stem from the variation in their halo mass evolution. Figure~\ref{fig:Vmax_Tvir} shows the time evolution of the maximum halo circular velocity measured using AHF and the corresponding virial temperature. The circular velocity of Halo 284 started to increase earlier than that of Halo 230, although the final circular velocities were similar for both halos. 
This difference led to earlier star formation and a higher star formation rate in Halo 284. \citet{2009MNRAS.399L.174O} showed that halos with a circular velocity above $12$\,km\,s$^{-1}$, i.e., $5\times 10^3$\,K can retain gas over the reionization epoch. The circular velocity of Halo 284 exceeds this threshold, but that of Halo 230 does not. In \citet{2009MNRAS.399L.174O}, halos below this threshold did not form stars at all, but our model forms stars. This may be due to a higher resolution of our simulations compared to theirs, and therefore, Halo 230 may also have locally reached a density high enough for star formation.
Indeed, the timing of the first star formation tends to delay as the resolution decreases (see Figure~\ref{fig:AnalysisData_Time-SFR}). 

Compared to the Hydro and Hydro Low models, the star formation rates of the low-resolution models were always lower than those of the high-resolution ones.
Consequently, the resulting stellar mass was higher for higher resolution models.
On the other hand, the half-mass radius of stars decreased with higher resolution. The stellar half-mass radius of Halo 284, the model with a larger $M_{\star}/M_{\rm vir}$, shrank more (from 178 to 36 pc) with a higher resolution compared to Halo 230 (from 118 to 68 pc). We also tested a case with even higher resolution ($2.37 M_{\odot}$ for gas particles), although the simulation was run only to $z=8.72$ ($t=0.568$\,Gyr). The comparison is shown in Appendix~\ref{sec:resolution}.

\begin{figure*}
	\begin{center}
        \includegraphics[width=1.0\linewidth]{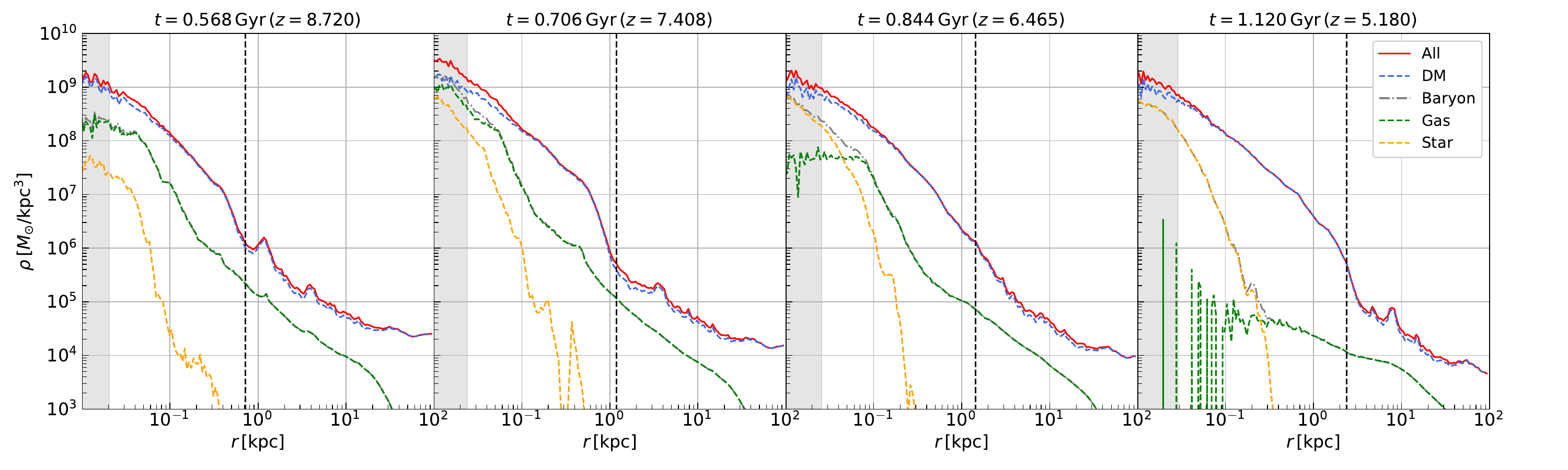}
	\end{center}
    \caption{Radial distribution of the density ($\rho$) of each component of the Halo 230 Hydro model from the early stage of star formation ($t=0.568$\,Gyr) to the end of star formation ($t=1.120$\,Gyr). The gray region shows the convergence radius obtained with eq. (\ref{equation:PowerRadius}) and the black dashed vertical line indicates the virial radius.
    Alt text: Three-panel figure with four lines for each panel. Red, blue, gray, green, and yellow indicate total, DM, baryon, gas, and star components. The density decreases with radius. The panels are for 0.568, 0.706, 0.844, and 1.120 gigayears.}
   \label{fig:RadialDensityTempSNSFCounts_List}
\end{figure*}

\begin{figure*}
	\begin{center}
        \includegraphics[width=1.0\linewidth]{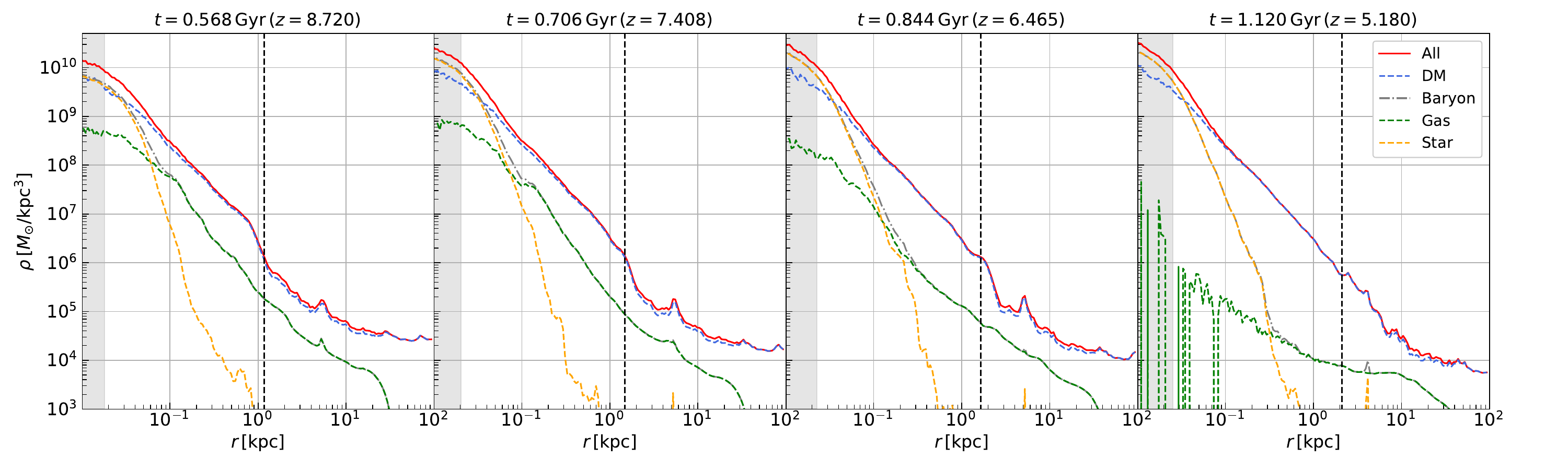}
	\end{center}
    \caption{Same as \ref{fig:RadialDensityTempSNSFCounts_List} but for Halo 284.
    Alt text: Three-panel figure with four lines for each panel. Red, blue, gray, green, and yellow indicate total, DM, baryon, gas, and star components. The density decreases with radius. The panels are for 0.568, 0.706, 0.844, and 1.120 gigayears.}
\label{fig:RadialDensityTempSNSFCounts_List.h284}
\end{figure*}

We also followed the time evolution of the density distribution of each component.
Figures ~\ref{fig:RadialDensityTempSNSFCounts_List} and \ref{fig:RadialDensityTempSNSFCounts_List.h284} present the time evolution of the radial profiles of each component of Halo 230 and Halo 284 Hydro models, respectively. 
Once stars formed, the H\textsc{ii} region appeared, and a fraction of the gas was heated to $10^4$\,K. Supernovae also occurred, but their energy was insufficient to stop star formation at this moment. The gas density inside the galaxies was maintained to be high enough to continue star formation. 
After the reionization epoch ($z=8.5$, $t=0.586$\,Gyr), the gas in the outer region of the galaxies was heated to more than the escape velocity, and the gas component started to decrease. 

In Halo 230, the gas dominated the stars in the innermost region until $t=0.706$\,Gyr (before the reionization in our simulation). At $t=0.844$\,Gyr, the gas in the central region started to decrease, and the stars became dominant. At $t=1.120$\,Gyr, the gas was depleted, and star formation ended. 

In Halo 284, the star formation started earlier than in Halo 230, and the stars dominated both gas and DM in the central region at $t=0.568$\,Gyr, which is before the reionization epoch. Since the central gas density was high enough at this moment, Halo 284 continued star formation after stars dominated the central region. The reionization contributed to evaporating the gas, but star formation continued even after the reionization epoch. The gas of the central region was finally depleted at $t=1.120$\,Gyr.

We measured the properties of the halo and stellar distribution at $z=0.5$. The results are summarized in Table~\ref{table:HaloProperty}. The halo viral mass and radius were calculated with the center determined using the method described in Section \ref{section:HaloCenterSelection}. 
The ratio of stellar to halo virial masses ($M_{\star}/M_{\rm vir}$) was $1.65\times10^{-4}$ for the Halo 230 Hydro model, which is similar to the satellites of the Milky Way. In contrast, $M_{\star}/M_{\rm vir}$ of Halo 284 ($3.05\times 10^{-3}$ for Hydro) was an order of magnitude higher than that of Halo 230. This is a typical value for galaxies that are more massive than the satellites of the Milky Way \citep{Lazar_2020}.

\begin{table*}
 \caption{Properties of the dwarf galaxies at $z=0.5$.}
 \label{table:HaloProperty}
 \centering
  \begin{tabular}{ccccccccc}
   \hline
    Halo Number & Model & $M_{\mathrm{vir}}\,[M_{\odot}]$ & $R_{\mathrm{vir}}\,[\mathrm{kpc}]$ & $M_{\star}\,[M_{\odot}]$ & $R_{1/2}^{\mathrm{3D}}\,[\mathrm{pc}]$ & $\bar{R}_{1/2}^{\mathrm{2D}}\,[\mathrm{pc}]$ & $V_{\mathrm{max}}\,[\mathrm{km}/\mathrm{s}]$ &
    $M_{\star}/M_{\mathrm {vir}}$\\
   \hline \hline
   230 & DMO & $9.34 \times 10^8$ & $17.2$ & - & - & - & $18.5$ & - \\
    & Hydro & $7.45 \times 10^8$ & $15.9$ & $1.23\times 10^5$ & $68.3$ & $52.7$ & $16.5$ & $1.65\times10^{-4}$\\
    & Hydro Low & $7.66 \times 10^8$ & $16.1$ & $3.45\times 10^4$ & $117.7$ & $90.7$ & $16.7$ & $4.50\times10^{-5}$ \\
   284 & DMO & $8.07 \times 10^8$ & $16.3$ & - & - & - & $18.1$ & -\\
    & Hydro & $6.95 \times 10^8$ & $15.5$ & $2.12\times 10^6$ & $36.4$ & $28.4$ & $19.3$ & $3.18\times10^{-3}$\\
    & Hydro Low & $6.94 \times 10^8$ & $15.5$ & $8.97\times 10^5$ & $178.2$ & $137.9$ & $17.0$ & $1.29\times10^{-3}$ \\
   \hline
  \end{tabular}

  From the third onward: (1) halo virial mass, (2) halo virial radius, (3) stellar mass within $0.2R_\mathrm{vir}$, (4) three-dimensional half-stellar-mass radius, (5) projected half-stellar-mass radius, (6) maximum circular velocity, and (7) $M_{\star}-M_{\mathrm{vir}}$.
\end{table*}

\subsection{DM density profile}
\label{section:DMRadialDensity}

\begin{figure*}
	\begin{center}
		\includegraphics[width=0.95\columnwidth]{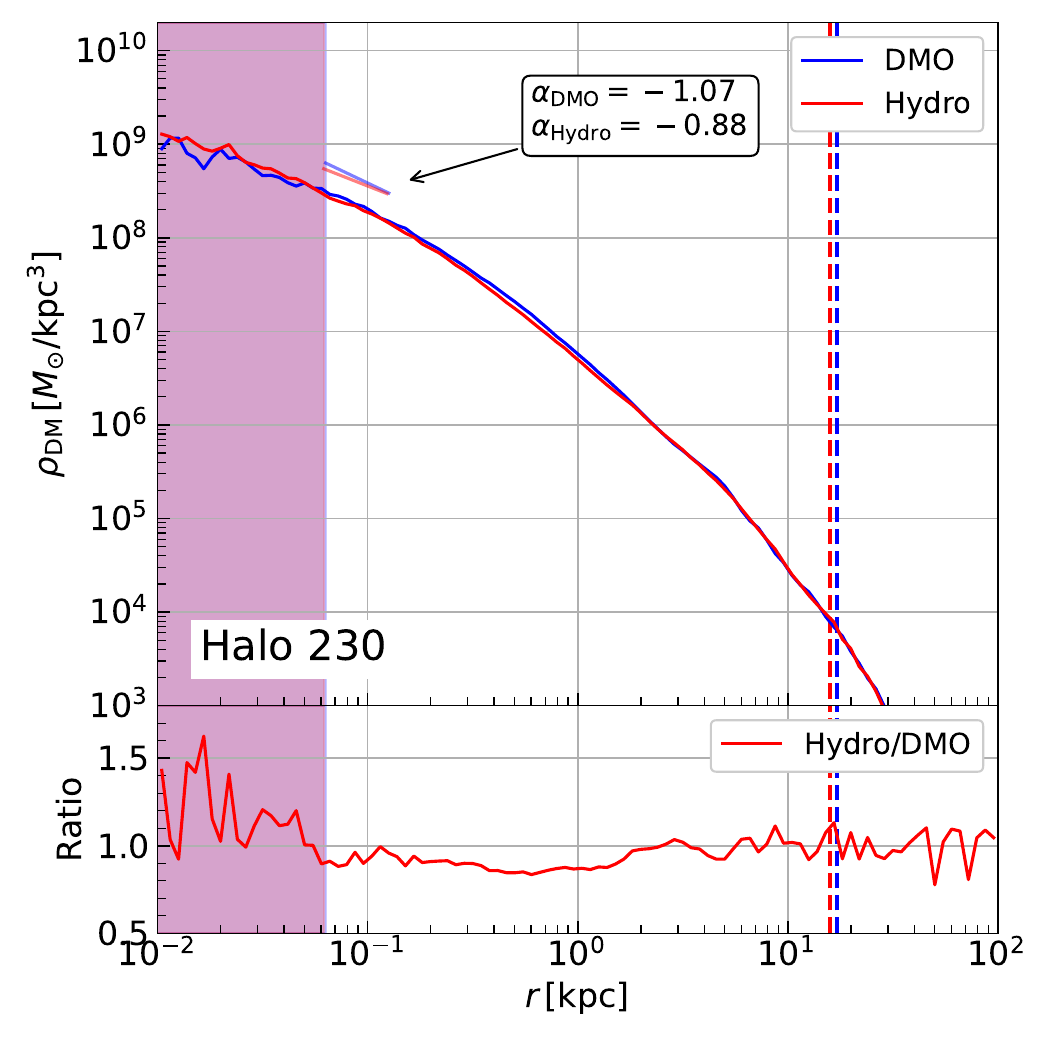}
        \includegraphics[width=0.95\columnwidth]{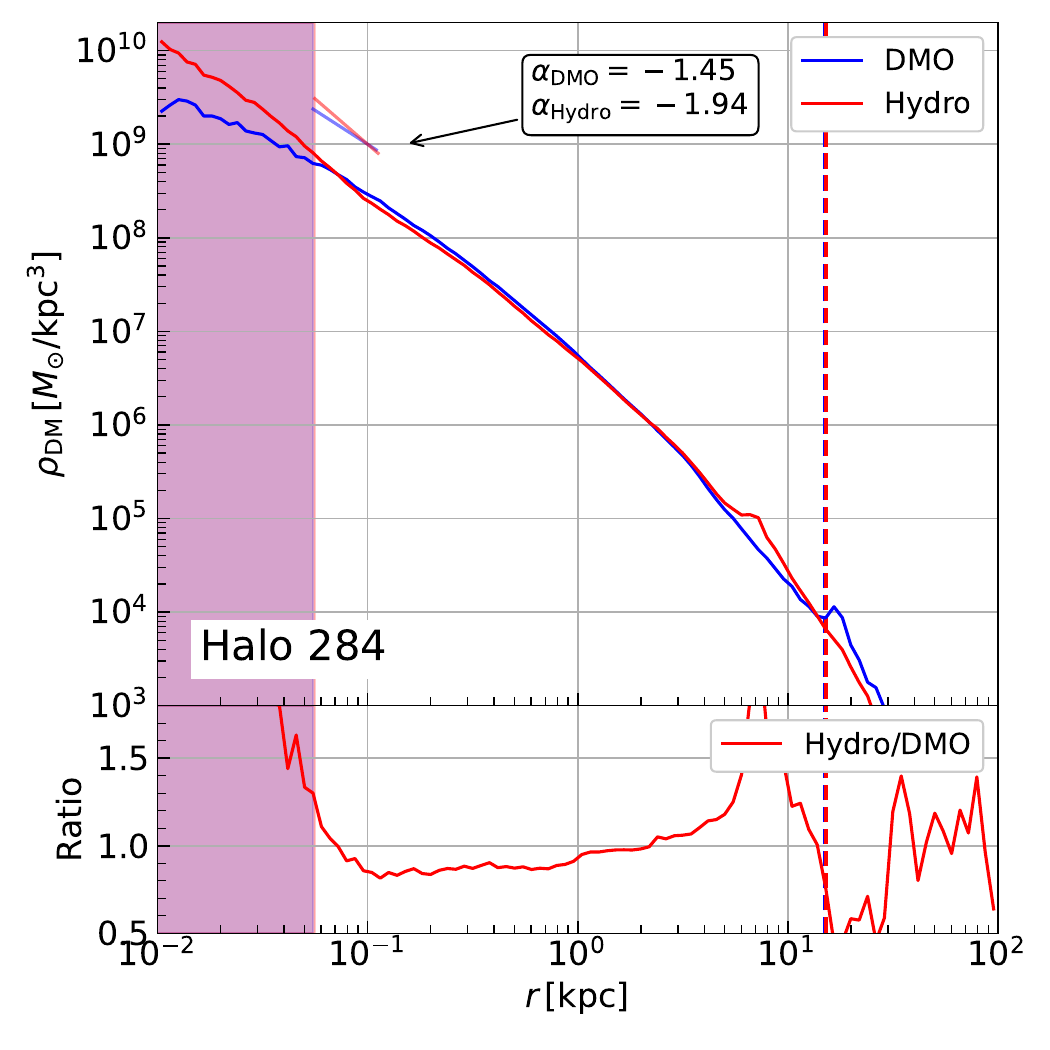}
	\end{center}
    \caption{Top panel: density profiles of the DM halo of the Hydro and DMO models (top panel) for Halo 230 (left) and Halo 284 (right) at $t=8.56\,\mathrm{Gyr}$ ($z=0.50$) and $t=8.29\,\mathrm{Gyr}$ ($z=0.54$), respectively. The colored regions indicate the convergence radii ($R_\mathrm{conv}$). The vertical dashed lines indicate the virial radii. The red and blue lines above the profiles indicate the slopes obtained from power-law fitting to $1-2\,R_\mathrm{conv}$ for the DMO and Hydro models, respectively. Bottom panels: the ratio of Hydro and DMO models. Alt text: Two-panel figures that include two panels. Two-line figure (top panel) and one-line figure (bottom) for each. In the top panels, blue and red indicate DMO and Hydro models, respectively. The fitted powers are -1.07 and -0.88 for the DMO and Hydro models of Halo 230. They are -1.45 and -1.94 for the DMO and Hydro models of Halo 284. }
	\label{fig:DMRadialDensity}
\end{figure*}

Figure~\ref{fig:DMRadialDensity} presents the density profiles at $t=8.56$\,Gyr and $t=8.29$\,Gyr for Halo 230 and Halo 284 Hydro models, respectively. Because Halo 284 underwent a merger around the final snapshot, we adopted a snapshot taken before the merger. 
For comparison, the DM profiles of the DMO models are also shown in the figure, but we reduced the mass by 15.6\,\% as the baryon mass.
For Halo 230, the DM halo profile remained unchanged regardless of the baryonic components. In contrast, for Halo 284, the DM halo density increased significantly in the innermost region, along with the baryonic components.

We evaluated the power-law slope ($\alpha$) of the innermost region of the density profile. In this study, the power-law slope was defined to be between the convergence radius (eq.~\ref{equation:PowerRadius}) and twice that radius similar to \citet{2025arXiv251110582S_LYRA}. Previous studies \citep{Cintio_2014,Tollet_2016,Lazar_2020} used the power-law slope at 1--2\,\% of the virial radius at $z=0$ ($R_{\mathrm{vir}}(z=0)$). However, in our simulations, 1--2\,\% of $R_{\mathrm{vir}}(z=0)$ was too far from the central region (see Appendix~\ref{sec:fitting}).
The density profiles and fitted $\alpha$ at $z=0.5$ are shown in Figure~\ref{fig:DMRadialDensity}. We obtained $\alpha = -0.88$ and $-1.07$ for the Hydro and DMO models of Halo 230, respectively. As can also be seen from the density profiles, the powers did not change much with the existence of the baryonic components. 
For Halo 284, we obtained $\alpha=-1.94$ and $-1.45$ for the Hydro and DMO models, respectively. Thus, Halo 284 is more cuspy with baryonic components. This is due to the dense stellar components in the central region of Halo 284. As seen in Figure~\ref{fig:RadialDensityTempSNSFCounts_List.h284}, Halo 284 is baryon-dominant in the central region. This made the central potential deeper, and as a result, the DM density also increased.

In both halos, we did not find any clear cores of DM distribution as was pointed out in previous simulations (e.g., \cite{Tollet_2016,Fitts_2017,Wheeler_2019,Orkney_2021}). This could be because of the very small halo mass compared to previous studies. For example, the halo mass in the study of \citet{Wheeler_2019} is close to $10^{10}M_{\odot}$. Massive halos have deeper potentials and therefore can cause multiple starbursts, which are necessary to make the transition from cusp to core. 
\citet{2025MNRAS.536..314M} pointed out that the amount of star formation that occurs earlier than the reionization epoch deepens the DM halo potential, as seen in our Halo 284. They also showed that the star formation occurring later than the reionization epoch is crucial for creating a core. However, both of our models ended star formation just after the reionization epoch.

\subsection{Time evolution of the power-law slope}

\begin{figure*}
	\begin{center}
	\includegraphics[width=0.95\columnwidth]{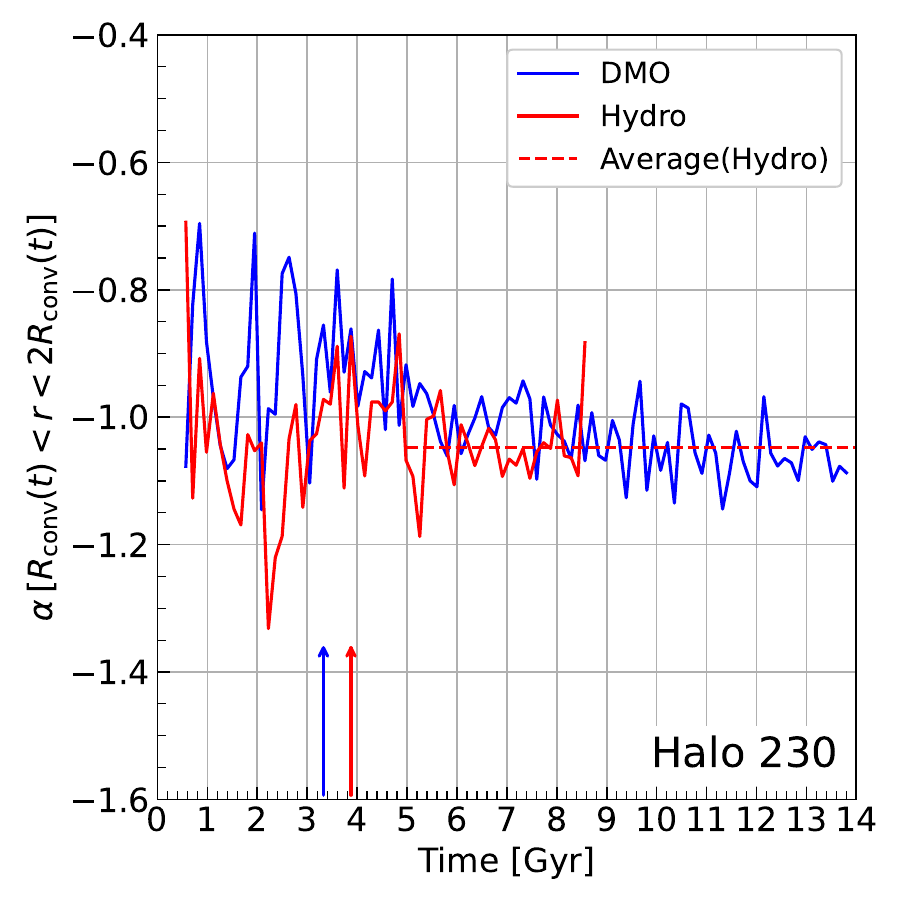}
	\includegraphics[width=0.95\columnwidth]{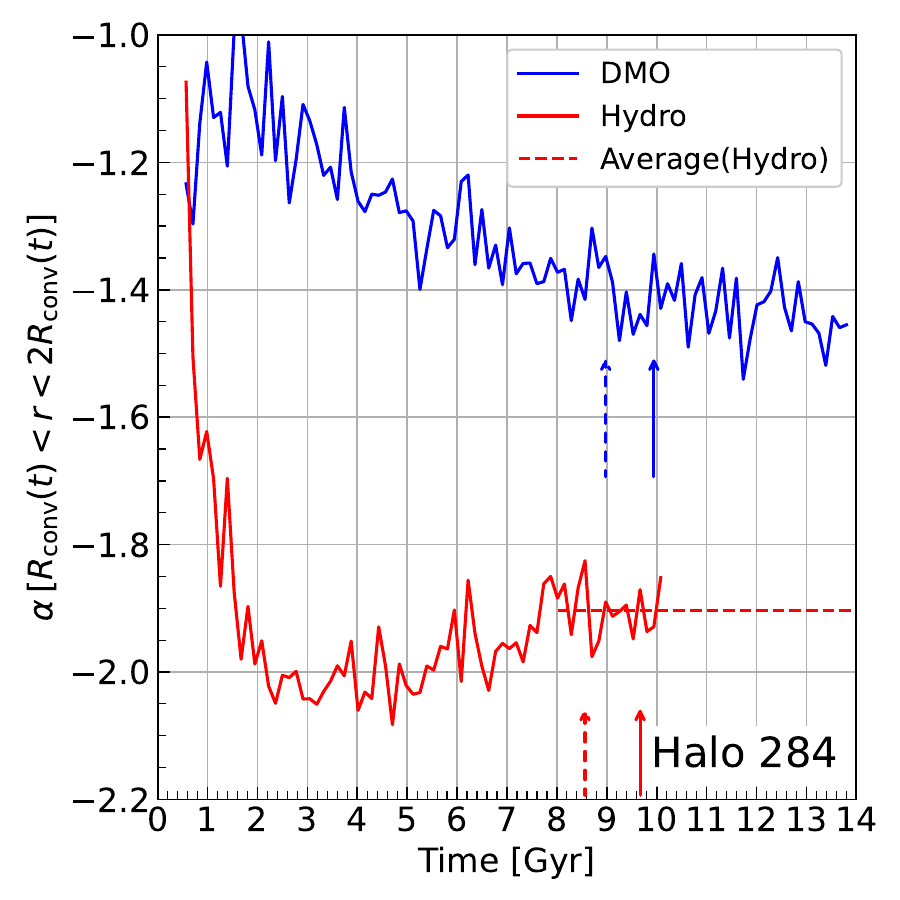}
	\end{center}
	\caption{Time evolution of the slopes obtained from the power-law fitting to $1-2\,R_{\mathrm{conv}}(t)$ for Halo 230 (left) and Halo 284 (right) models. The red and blue curves indicate the Hydro and DMO models, respectively. The red dashed line indicates the slope averaged in 5.00--8.56 Gyr for Halo 230 ($\alpha = -1.01$) and 8.01--10.1 Gyr for Halo 284 ($\alpha = -1.90$). The dashed and solid arrows indicate the time of close encounters and mergers, respectively.
    Alt text: Two-line figures. In the left panel, blue and red arrows at 3.2 and 4 gigayears. In the right panel, blue dashed and full arrows are at 9 and 11 gigayears. Red dashed and full arrows are at 8.5 and 9.5 gigayears. }
	\label{fig:SlopeHistory_z0}
\end{figure*}

We investigated the time evolution of the power-law slope ($\alpha$) because it may evolve with time, as shown in the study by \citet{Tollet_2016}. The results are shown in Figure~\ref{fig:SlopeHistory_z0}.
After the last merger event at approximately 3--4\,Gyr, the power-law slope did not change significantly for the DMO and Hydro models. Averaging the slope of $t=5$--8.5\,Gyr, we defined the final slope as $\alpha = -1.01\pm0.01$ for the Hydro model.

In contrast, the power-law slope of the Hydro model of Halo 284 largely decreased in the first 2\,Gyr, in which the star formation occurred. During this period, the power-law slopes of the DMO and Hydro models were largely distinct, and the power of the Hydro model reached $-2.0$. 
The power-law slope of the DMO model gradually decreased from $-1.1$ to $-1.4$ with time. This is due to the increase of the convergence radius with time (see eq.~\ref{equation:PowerRadius}). This halo experienced a merger at approximately 10\,Gyr. After that, the power was almost constant. On the other hand, the power-law slope of the Hydro model increased from $-2.0$ to $-1.9$ with time. Compared to the two halos, Halo 284 always had a steeper slope than Halo 230, even for the DMO models.  
Averaging the slope of $t=8$--10\,Gyr, we obtained $\alpha = -1.90\pm0.01$ for the Hydro Model.
The decrease in $\alpha$ of the Halo 284 DMO model may be due to the change in the convergence radius with time.

\section{Discussion: comparison with observations and previous simulations}
\subsection{Dark matter distribution}

\begin{figure*}
	\begin{center}
	\includegraphics[width=0.95\linewidth]{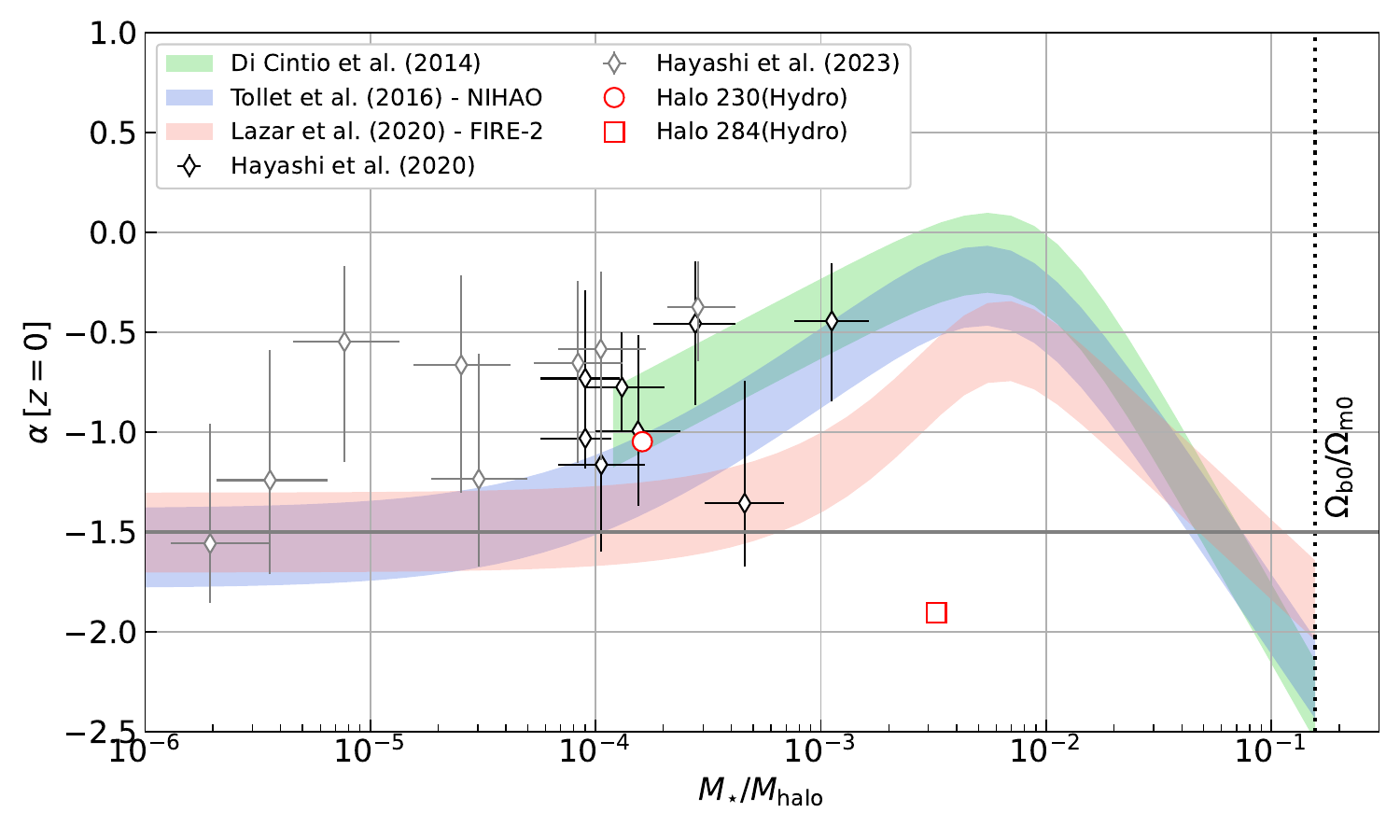}
	\end{center}
	\caption{Relationship between the stellar-to-halo mass ratio ($M_{\star}/M_{\mathrm{halo}}$) and the inner power-law slope of DM halo ($\alpha$). The green, blue, and red ranges are the relationships obtained from the simulation results of \citet{Cintio_2014}, \citet{Tollet_2016}, and \citet{Lazar_2020}, respectively, and the vertical width of the ranges for all is $\Delta \alpha = 0.2$. The diamonds with error bars represent the observations of dwarf galaxies \citep{Hayashi_2020,2023ApJ...953..185H}. The red circle and square indicate the results of the Hydro models for Halo 230 and Halo 284, respectively.  
    Note that $\alpha$ of the Hydro models is the average value of earlier times (see Fig.~\ref{fig:SlopeHistory_z0}). Alt text: Scatter plot with three shaded regions. }
	\label{fig:MstarMhalo_Alpha}
\end{figure*}

Figure~\ref{fig:MstarMhalo_Alpha} shows the relationship between the stellar-to-halo mass ratio ($M_{\star}/M_{\mathrm{halo}}$) and $\alpha$ obtained from our simulations, observed dwarf galaxies \citep{Hayashi_2020,2023ApJ...953..185H}, and previous simulations \citep{Cintio_2014,Tollet_2016,Lazar_2020}. 
For the previous simulations, we fitted functions to their results and plotted the regions shown in the figure. The details of this process are described in Appendix~\ref{sec:previous_sim}. For the observations, fittings were performed using the Hernquist profile in ellipsoidal geometry \citep{Hayashi_2020,2023ApJ...953..185H}, whereas simulations assumed a spherical model.

Halo 230 is consistent with those obtained from observations and previous simulations. 
On the other hand, the value of $\alpha$ for Halo 284 is significantly smaller than those obtained from previous simulations, and $M_{\star}/M_{\mathrm{halo}}$ is much higher than those of the satellites of the Milky Way. 
We note that the dwarf galaxies with $M_{\star}/M_{\rm halo}\sim 3\times 10^{-3}$ in previous simulations typically have a DM halo with a mass of  $10^{10}$--$10^{11} M_{\odot}$, whereas Halo 284 has $\sim 10^9M_{\odot}$.

In the sample of \citet{2025arXiv251110582S_LYRA}, which is a simulation with the mass resolution of $4 M_{\odot}$, there was a dwarf galaxy which shows a stellar-to-halo mass ratio similar to that of Halo 284. It (their Halo A) also showed a high stellar-to-halo mass ratio of order of $10^{-3}$, similar to our Halo 284. Their Halo A showed a deeper cusp in the hydrodynamical simulation than in the DMO simulation. However, the stellar distribution in their simulation had a mass of $\sim 10^7 M_{\odot}$ and a half-mass radius of $\sim 300$\,pc. They are an order of magnitude larger than our Halo 284. 

In the samples in \citet{2021MNRAS.504.3509O_EDGE}, they showed that the dark matter density in the central region with baryons decreased compared to DMO simulations due to the baryon physics, and the degree of the decrease becomes larger in more massive halos. On the other hand, we did not see any decrease in the dark matter density, although our sample size is only two. However, their halo masses are $>1.3\times 10^9 M_{\odot}$, which is more massive than our simulations ($<10^9 M_{\odot}$). 

\begin{figure*}
	\begin{center}
	    \includegraphics[width=0.8\linewidth]{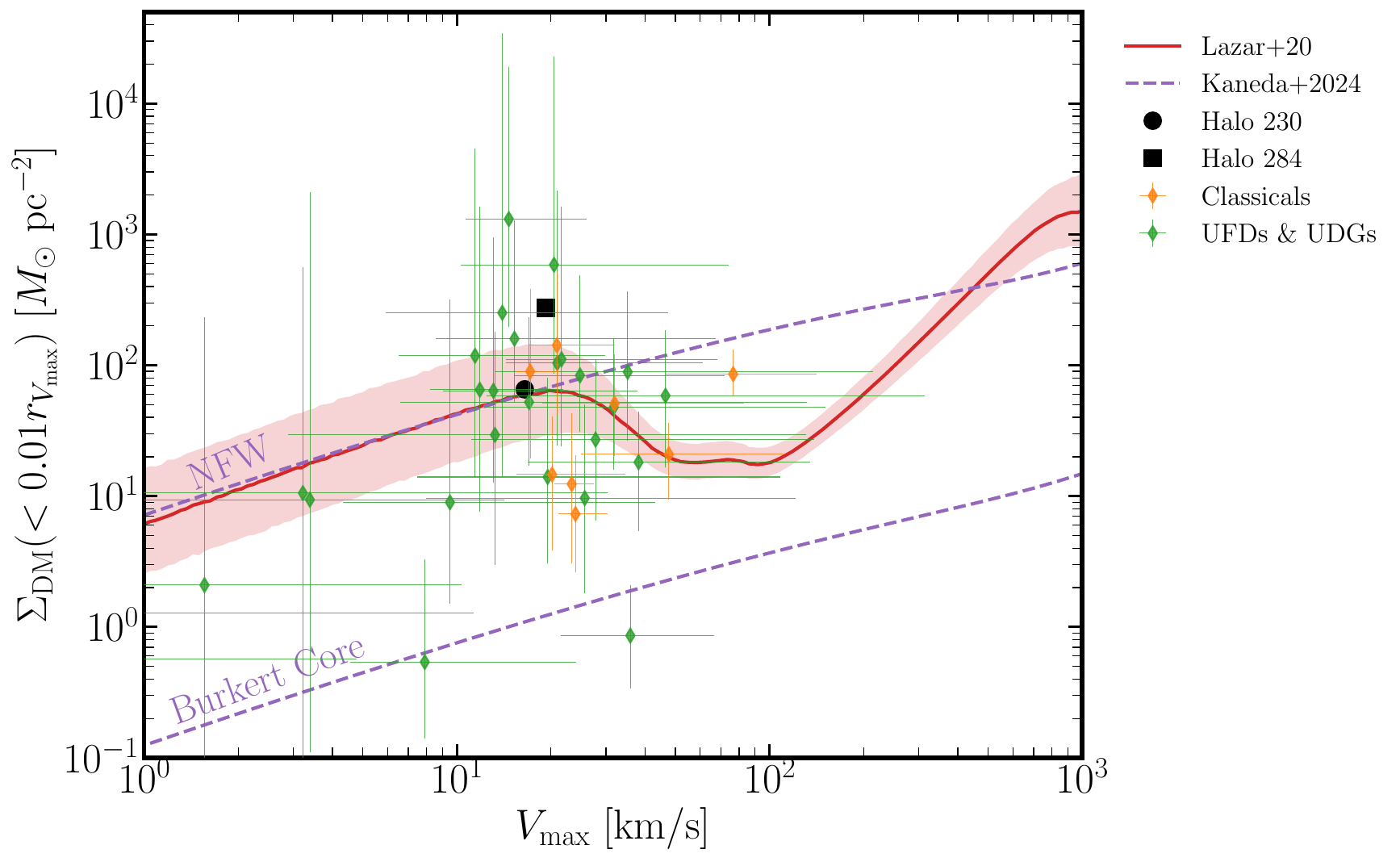}
	\end{center}
	\caption{Central surface density of the DM halo within 1\,\% of the radius of the maximum circular velocity, $\Sigma _{\rm DM}(<0.01r_{\rm V_{max}}) $ as a function of the maximum circular velocity, $V_{\rm max}$. The diamonds with error bars show the observed dwarfs; the orange diamonds indicate classical dwarfs associated with the Milky Way \citep{Hayashi_2020}, whereas the green ones indicate ultra-faint dwarfs and ultra-diffuse galaxies \citep{2023ApJ...953..185H}. The red line indicates the result of \citet{Lazar_2020}. The red shaded region shows the $1\sigma$ dispersion. The two purple dashed lines represent the predictions from \citet{2024PASJ...76.1026K} for the NFW (cuspy) and Burkert (cored) profiles, respectively.
    Alt text: Scatter plot with three lines and a shaded region. 
    }
	\label{fig:Comp_Kaneda}
\end{figure*}

\citet{2024PASJ...76.1026K} proposed a condition for the transition from cusp to core based on the DM halo density in the central region rather than the stellar-to-halo mass ratio. \citet{2025arXiv250722155H} showed the relation between the maximum rotation speed, which indicates the halo mass, and the DM halo density in the central region of the observed dwarf galaxies. The observation aligns more closely with the simulation than with the stellar-to-halo-mass ratio. We plotted the same relationship between our simulated galaxies and observations \citep{Hayashi_2020,2023ApJ...953..185H} and previous simulations \citep{Lazar_2020} in Figure~\ref{fig:Comp_Kaneda}. In this plot, some ultra-faint and ultra-diffuse galaxies show dark matter surface densities as high as those of Halo 284.

\subsection{Stellar distribution}

\begin{figure}
	\begin{center}
	\includegraphics[width=0.95\columnwidth]{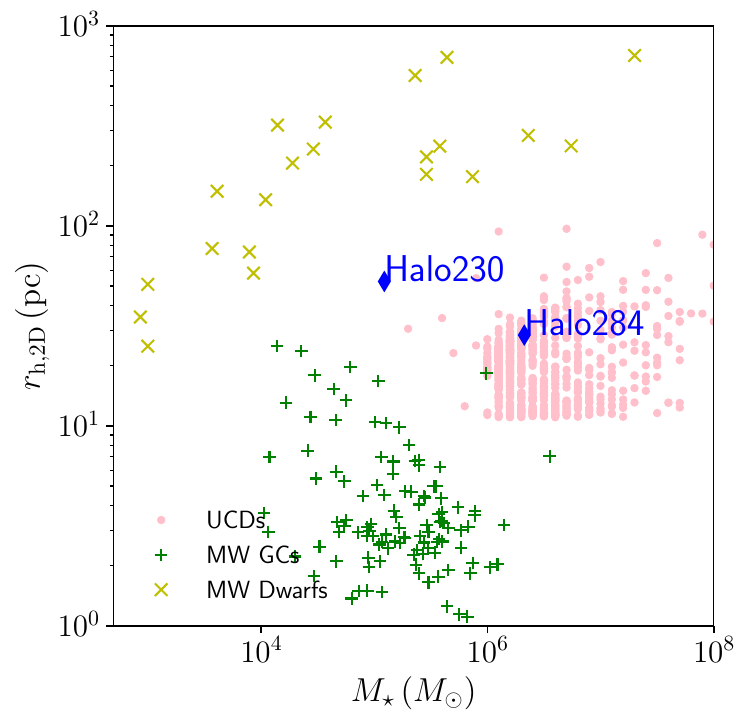}
	\end{center}
	\caption{Relation between the mass of galaxies and their projected half-mass radius. The yellow crosses and green pluses indicate dwarf galaxies \citep{2012AJ....144....4M} in the local group and globular clusters in the Milky Way \citep{2018MNRAS.478.1520B}, respectively. The pink dots indicate ultra-compact dwarfs \citep{2020ApJS..250...17L}, and the blue diamonds show our simulated galaxies. Alt text: Scatter plot for three types of stellar systems and simulated galaxies. 
    }
	\label{fig:UCDs}
\end{figure}

We also compared the stellar mass and half-mass radius with those of the Milky Way globular clusters and dwarf galaxies, along with ultra compact dwarfs (UCDs), in Figure~\ref{fig:UCDs}. 
Halo 230 was close to the satellites of the Milky Way, but had a slightly smaller half-mass radius. This mass-radius relation is consistent with 
Halo 284 was located in the region of UCDs, rather than the satellites of the Milky Way. 
In the halos of \citet{2025arXiv251110582S_LYRA}, Halo D has the most compact stellar distribution in their samples. The stellar mass of their Halo D was $1.2\times 10^6 M_{\odot}$, which is similar to our Halo 284, but their half-mass radius of 120\,pc was larger than that of ours (28\,pc). Different from their Halo A, which shows a steeper cusp than the DMO simulation, their Halo D did not show such a trend. In addition, the stellar-to-halo mass ratio of their Halo D was $<10^{-3}$, which is lower than that of Halo 284.

As shown in Figure~\ref{fig:Vmax_Tvir}, the circular velocity of Halo 284 increased much faster than in Halo 230, and more stars were formed before the reionization epoch. Such an early evolution may have resulted in the formation of a UCD-like galaxy.
However, the results of this study were obtained for only two halos, and therefore, more robust results would require similar simulations of other halos to see if there is any variation among galaxies.

\section{Summary}
In this study, we performed cosmological zoom-in simulations of dwarf galaxies of DM halos with masses of $\sim 10^9M_{\odot}$ at $z=0$. The mass resolution was $19$ and $102 M_{\odot}$ for gas and DM particles, respectively.
The evolution of halo density profiles was investigated. 
One of our two dwarf galaxies (Halo 230) was similar to ultra-faint or classical dwarfs with a stellar-to-halo-mass ratio of $\sim 10^{-4}$. The DM density profile was not altered from that in the DM-only simulation, and the powers of the innermost slope were $-1.07$ and $-0.88$ for the DM-only and the Hydro simulations, respectively. Once the halo was formed, its profile remained unchanged over time.
The other (Halo 284) had a larger stellar-to-halo mass ratio of $>10^{-3}$, which is typical for galaxies with halos that are an order of magnitude more massive than those in our samples. This galaxy was more compact compared to the satellites of the Milky Way and as compact as UCDs. The power of the DM halo profile with baryons was steeper ($-1.94$) than that of the DM-only simulation ($-1.45$). We did not find any cusp-to-core transition in our samples. 

We also investigated the dependence of the galaxy structures on the resolution. In our low-resolution runs (Hydro Low), the stellar distribution of Halo 284 was more extended than that of Halo 230. In our high-resolution runs (Hydro), the half-mass radius of Halo 284 (28.4\,pc) became smaller than that of Halo 230 (52.7\,pc). The stellar masses were $2.12\times 10^6$ and $1.23\times10^5 M_{\odot}$ for Halo 284 and 230, respectively. The mass and density of Halo 284 are comparable to those of UCDs, and we found it only in simulations with sufficiently high resolution. We further performed a higher-resolution simulation for Halo 284, although it is still in the early phase of star formation due to the calculation cost. This model maintained its compact structure even after starting supernova feedback. We, therefore, considered that the simulation results had converged. 

The difference between these two dwarfs may come from the halo mass at the reionization epoch. Although the final halo mass is similar for both halos, the halo with a higher stellar mass (Halo 284) grows earlier than the other halo (Halo 230). The early growth of the halo mass before reionization makes the DM potential deep enough for the halo to continue star formation after reionization. Thus, the early evolution of the DM halo may be a necessary condition for the formation of UCDs.

%

\begin{ack}
Simulations and analyses in this study were carried out on Cray XC50 and analysis servers at the Center for Computational Astrophysics, National Astronomical Observatory of Japan, and Fujitsu PRIMERGY CX400M1/CX2550M5 (Oakbridge-CX) at the Supercomputer Division of the Information Technology Center, The University of Tokyo.
We thank Kohei Hayashi for providing the observational data of dwarf galaxies. We would like to thank Editage (www.editage.jp) for English language editing.
\end{ack}

\section*{Funding}
 This research was supported by KAKENHI Grant Numbers 23K22530, 21K03614, 21K03633, 22KJ0157, 22K03688, 24K07095, 25K01046, and 25H00664.

\section*{Data availability} 
The simulation data will be provided upon request.

\appendix 

\section{Comparison with a higher resolution simulation}\label{sec:resolution}

In order to see the effect of the resolution, we performed an additional higher resolution simulation only for Halo 284 (model Halo 284 Hydro High). The resolutions for gas and DM particles are 2.37 and $12.8 M_{\odot}$, respectively. The gravitational softening lengths are 2.84 and 4.50\,pc for the gas and DM. The time for the beginning of the star formation and the subsequent star formation rate of the Hydro High model were similar to those of the Hydro model, but the peak value of the star formation rate was slightly higher than that of the Hydro model (see Figure~\ref{fig:AnalysisData_Time-SFR}).

Due to the computational cost, we stopped the Hydro High simulation at $z=8.72$ ($t=0.568$\,Gyr). A comparison of the halo and stellar distributions of these models was done at $z=8.75$, and the results are summarized in Table~\ref{table:HaloPropertyHigh}. The stellar mass of the Hydro High model was approximately twice that of the Hydro model. 
The differences in the $M_{\star}/M_{\mathrm{vir}}$ ratio between these two models reflected those in their stellar masses.
The half-mass decreased when we increased the resolution from Hydro High to Hydro, but it did not decrease further. The half-mass density of the Hydro model ($5.8\times M_{\odot}$\,pc$^{-3}$) was much higher than that of the Hydro Low model ($0.14 M_{\odot}$\,pc$^{-3}$) but similar to that of the Hydro High model ($7.1\times M_{\odot}$\,pc$^{-3}$). 
Therefore, we consider that the results of our Hydro model converged, although the mass and density changed by a factor of two.

\begin{table}
 \caption{Same as Table~\ref{table:resolution} but for the highest resolution one}
 \label{table:resolution}
 \centering
  \begin{tabular}{lcccc}
   \hline
    Model & $m_{\mathrm{gas}}$ & $\epsilon_{\mathrm{gas}}$ &  $m_{\mathrm{DM, min}}$ & $\epsilon_{\mathrm{DM, min}}$\\
     & [$M_{\odot}$] & $[\mathrm{pc}]$ &  $[M_{\odot}]$ & $[\mathrm{pc}]$\\
   \hline \hline
   Hydro High & 2.37 & 2.84 & 12.8 & 4.50 \\
   \hline
  \end{tabular}
  
  From left to right: (1) gas-mass resolution ($m_{\rm gas}$), (2) gas softening length ($\epsilon_{\rm gas}$) (3) minimum dark-matter mass ($m_{\rm DM, min}$) (4) minimum dark-matter softening length ($\epsilon_{\rm DM, min}$). 
\end{table}

\begin{table*}
 \caption{Properties of the dwarf galaxies at $z=8.72$ ($t=0.568$\,Gyr). The columns are the same as in Table~\ref{table:HaloProperty}.}
 \label{table:HaloPropertyHigh}
 \centering
  \begin{tabular}{llccccccc}
   \hline
    Halo Number & Model & $M_{\mathrm{vir}}\,[M_{\odot}]$ & $R_{\mathrm{vir}}\,[\mathrm{kpc}]$ & $M_{\star}\,[M_{\odot}]$ & $R_{1/2}^{\mathrm{3D}}\,[\mathrm{pc}]$ & $\bar{R}_{1/2}^{\mathrm{2D}}\,[\mathrm{pc}]$ & $V_{\mathrm{max}}\,[\mathrm{km}/\mathrm{s}]$ & 
    $M_{\star}/M_{\mathrm {vir}}$\\
   \hline \hline
   284 
   & Hydro Low & $4.78 \times 10^7$ & $1.16$ & $1.98\times 10^5$ & $69.7$ & $55.4$ & $13.8$ & $4.14\times10^{-3}$\\
    & Hydro & $4.90 \times 10^7$ & $1.17$ & $8.22\times 10^5$ & $32.4$ & $25.2$ & $14.0$ &$1.68\times10^{-2}$\\
    & Hydro High & $5.14 \times 10^7$ & $1.19$ & $2.12\times 10^6$ & $41.5$ & $31.8$ & $17.0$ & $4.12\times10^{-2}$\\
   \hline
  \end{tabular}
\end{table*}

Because of the finite number of particles, the resolution of the simulations may affect the cusp of the innermost region of the DM halo. 
We estimated the convergence radius over which the radial density profile was considered to converge to higher-resolution simulations. Inside it, the DM density profile could become shallower due to two-body relaxation. The convergence radius was calculated using the criterion suggested in \citet{Power_2003}:
\begin{equation}
\frac{t_{\mathrm{relax}}(r)}{t} = \frac{\sqrt{200}}{8} \frac{N(<r)}{\ln{N(<r)}} \left(\frac{\bar{\rho}(<r)}{\rho_{\mathrm{crit}}}\right)^{-1/2} > 0.6,
\label{equation:PowerRadius}
\end{equation}
where $N(<r)$ and $\bar{\rho}(<r)$ are the number of particles within a sphere of radius $r$ and the average density in the sphere, respectively. For the critical density, $\rho_{\rm crit}=3H^2/8\pi G$, where $H$ and $G$ are the Hubble parameter and gravitational constant, respectively. If this criterion is satisfied, the relaxation time is longer than the age of the Universe, and therefore, the condition for the collisionless system is satisfied. 
We defined the minimum radius satisfying this criterion as the convergence radius, $R_{\rm conv}$.
Other definitions, such as 2\,\% of the viral radius, are also used (e.g., \cite{Schaller_2015,Tollet_2016}), and they are smaller than that of \citet{Power_2003}.

\begin{figure}
	\begin{center}
    \includegraphics[width=0.9\columnwidth]{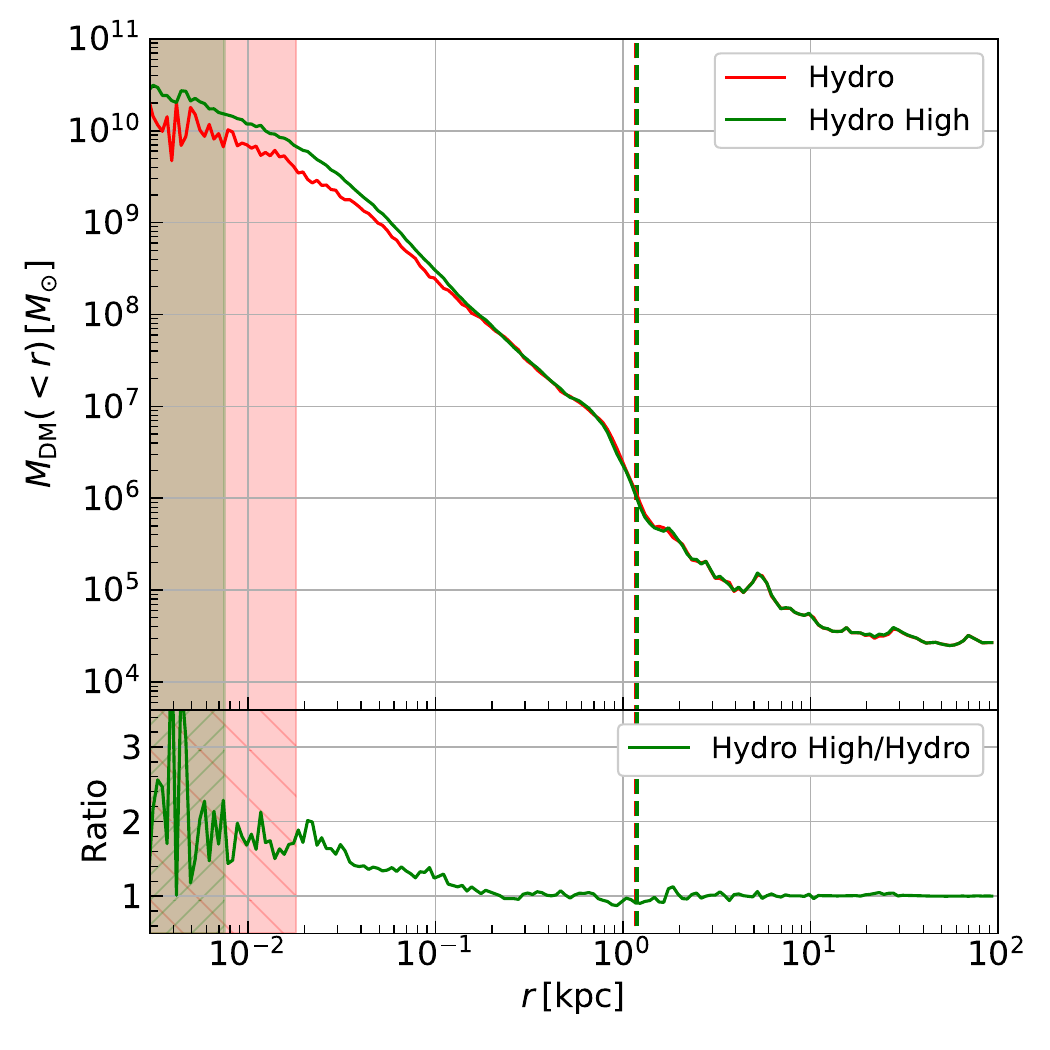}
	\end{center}
	\caption{DM density profiles (top) and their ratios (bottom) at $z=8.72$ in the Hydro and Hydro High models of Halo 284. The red and green shaded regions indicate the conversion radii. The vertical dashed lines indicate the virial radii. Alt text: Two-line figure with red (Hydro) and green (Hydro High) lines. }
	\label{fig:Comparision}
\end{figure}

Figure~\ref{fig:Comparision} shows the density profiles of the DM halo together with the convergence radii, $R_{\rm conv}$, obtained using equation (\ref{equation:PowerRadius}) for the Hydro and Hydro High models. As shown in the figure, the density profiles differ even outside the convergence radius. The density of the DM halo for the Hydro High model is higher than that of the Hydro model because of the higher stellar mass in the central region. Although the density of the Hydro High model is higher than that of the Hydro model, the power of the density profile just outside the convergence radius is similar for both models.

\begin{figure}
	\begin{center}
        \includegraphics[width=0.95\columnwidth]{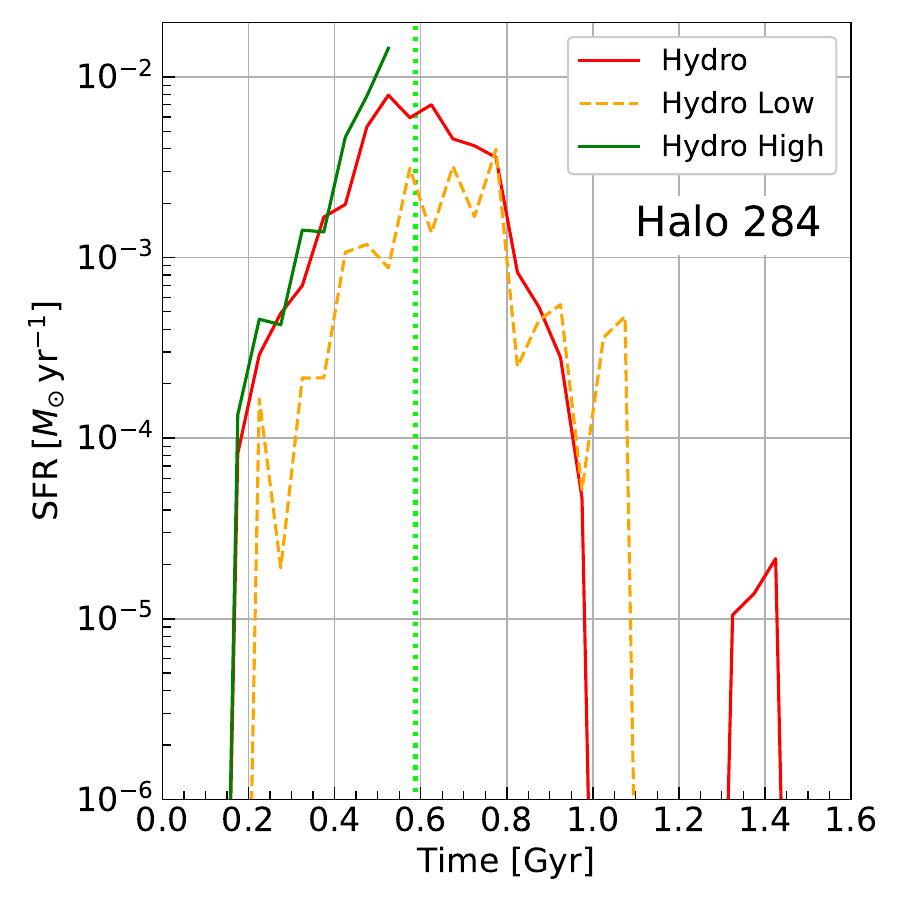}
	\end{center}
	\caption{Same as the bottom panel of Fig.~\ref{fig:AnalysisData_Time-SFR}, but with Hydro High. Alt text: Three-line figure. Red, orange, and green indicate Hydro, Hydro Low, and Hydro High models.}
	\label{fig:AnalysisData_Time-SFR_High}
\end{figure}

\begin{figure*}
	\begin{center}
	    \includegraphics[width=0.9\columnwidth]{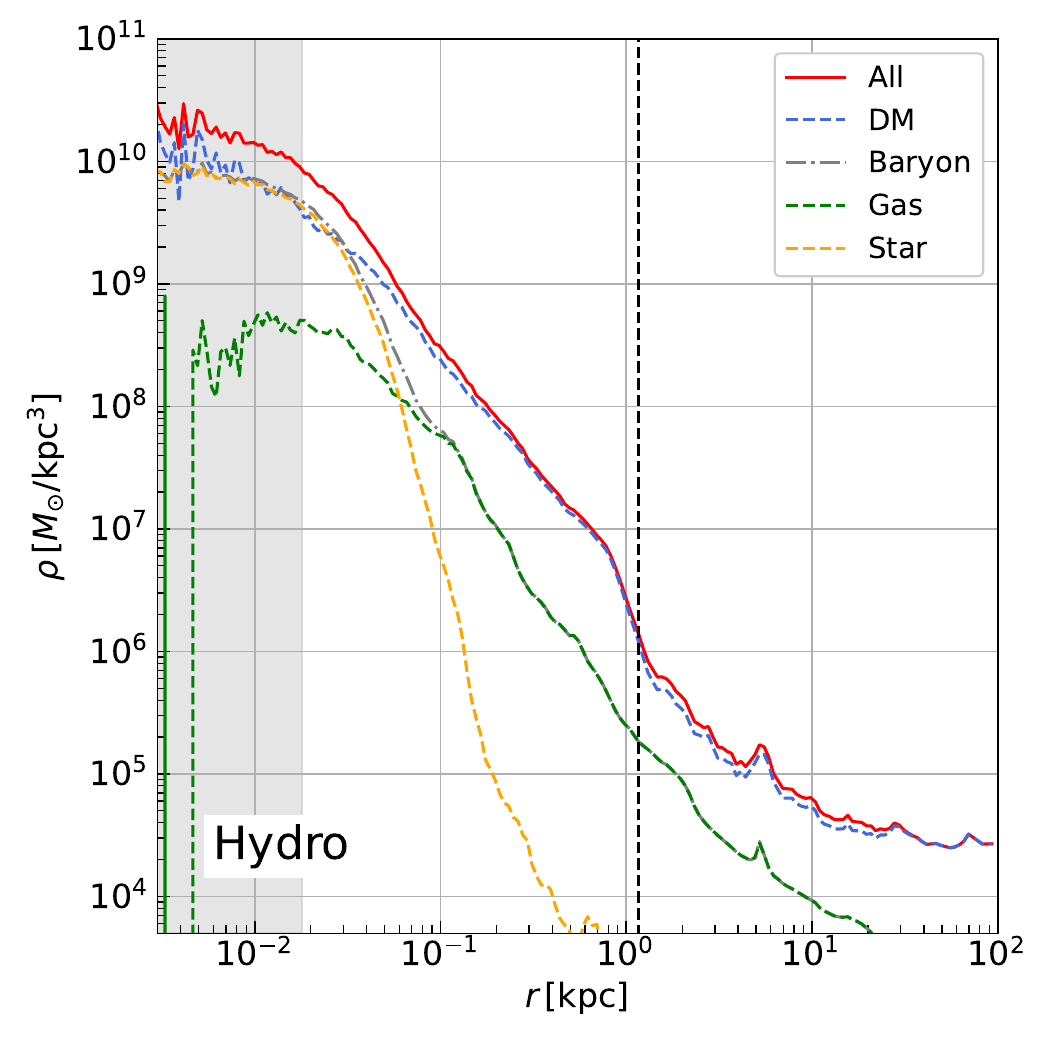}
        \includegraphics[width=0.9\columnwidth]{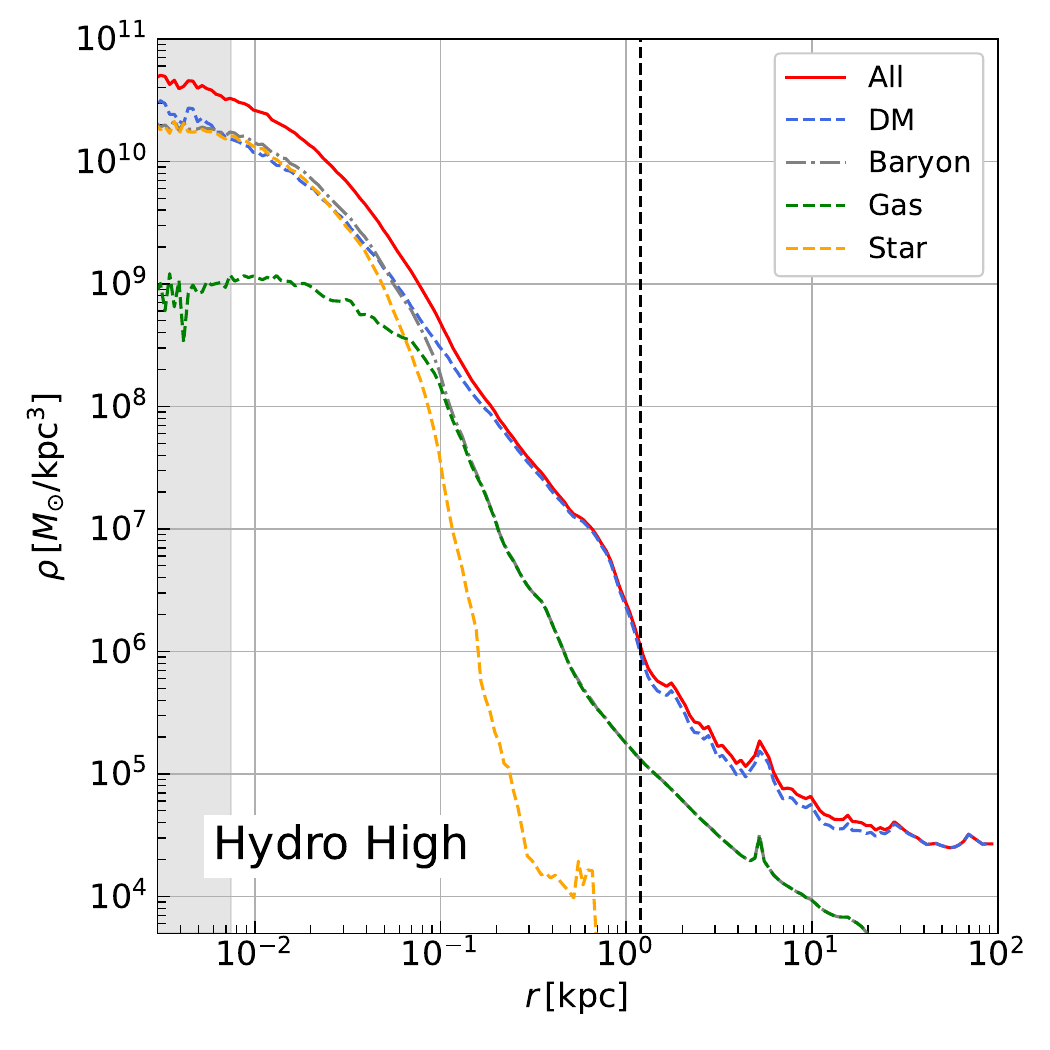}
	\end{center}
	\caption{Density profiles of gas, star, and DM at $z=8.72$ ($t=0.568$\,Gyr) for the Hydro (left) and Hydro High (right) models of Halo 284. The vertical black dashed line indicates the virial radius. Alt text: Two-panel figures with five lines for each. The colors are the same as Fig.~\ref{fig:RadialDensityTempSNSFCounts_List}.}
	\label{fig:lv13}
\end{figure*}

Figure~\ref{fig:lv13} shows the density profiles of the gas, star, and DM components of Halo 284 at $t=0.568$\,Gyr ($z=8.72$). Although the innermost region densities are higher in the Hydro High model, the characteristics of the three components are consistent: the DM and star densities are comparable, and the gas has a flat distribution in the central region. Thus, we consider that the Hydro simulation resolves the inner cusp in our simulations.

\section{Halo profile fitting}
\label{sec:fitting}
\begin{figure*}
	\begin{center}
		\includegraphics[width=0.9\columnwidth]{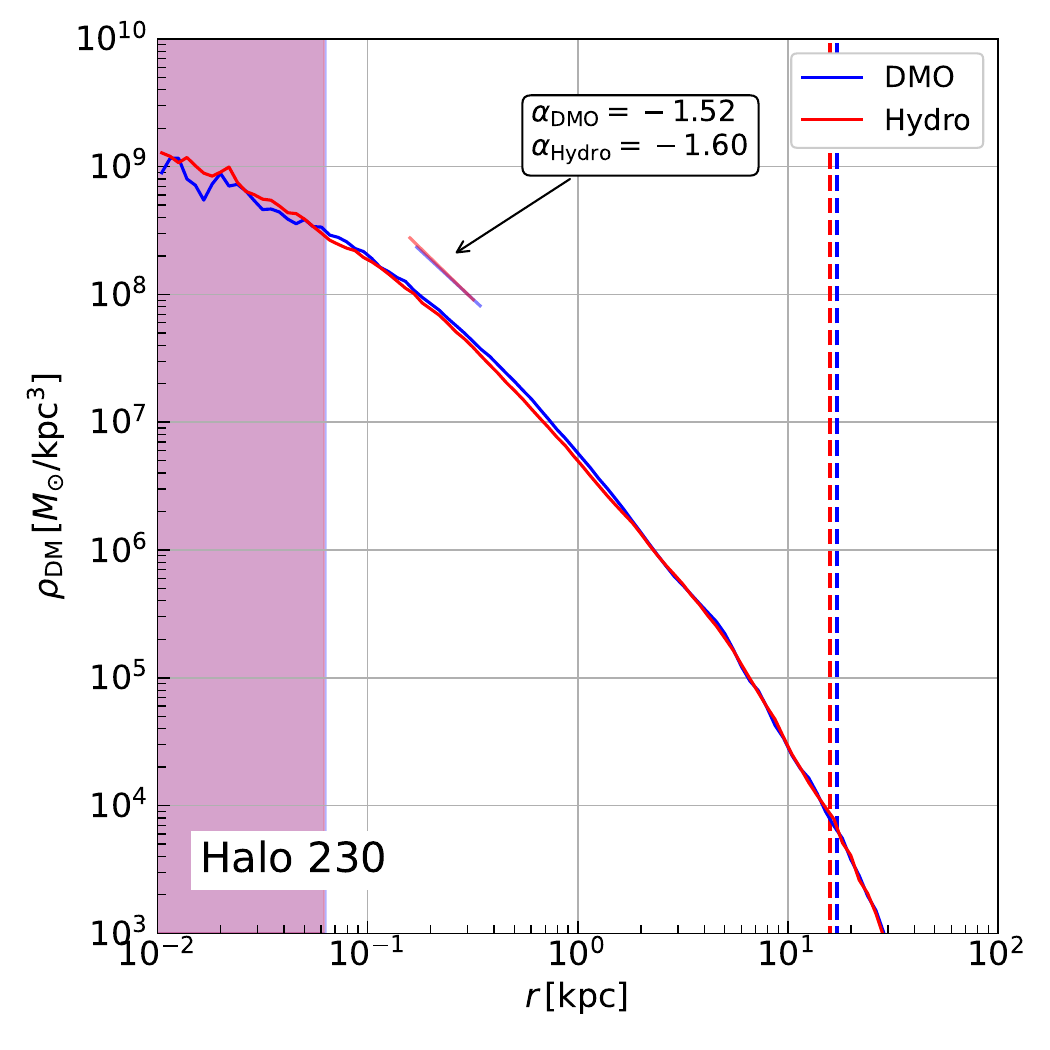}
        \includegraphics[width=0.9\columnwidth]{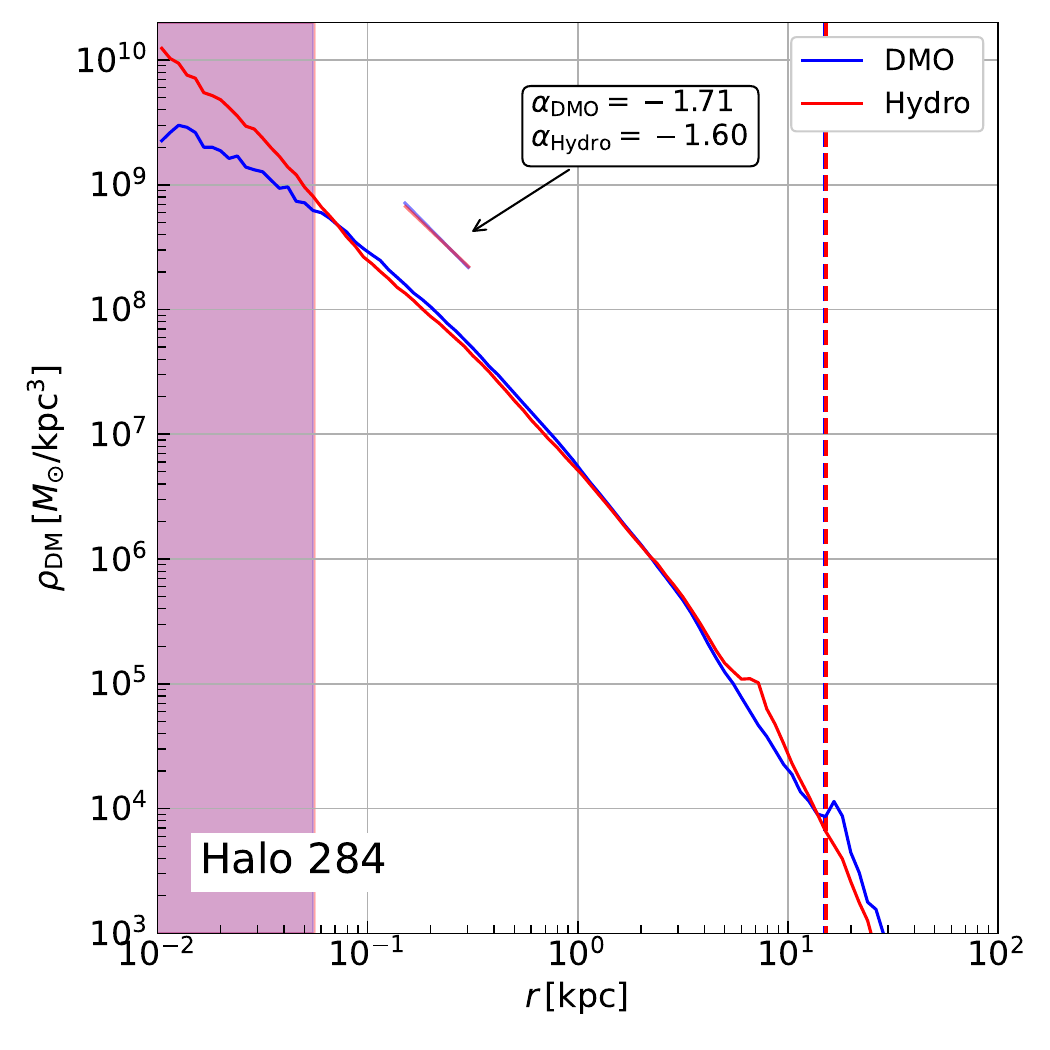}
	\end{center}
    \caption{Same as Figure~\ref{fig:DMRadialDensity} but fitted for 1--2\,\% virial radius at $z=0$, $R_{\rm vir} (z=0)$, for Halo 230 (left) and Halo 284 (right). The red and blue lines indicate the power-law slope and the range of 1--2\,\%$R_{\rm vir} (z=0)$. Alt text: Two-panel plot with two lines for each. Blue and Red indicate DMO and Hydro models, respectively. The fitted powers are -1.52 and -1.60 for DMO and Hydro, respectively, in Halo 230. They are -1.71 and -1.68 for DMO and Hydro, respectively, in Halo 284.}
	\label{fig:DMRadialDensity2}
\end{figure*}

In this study, we measured the power-law slope, $\alpha$, at 1--2\,$R_{\rm conv}$, which was the innermost region in our simulations similar to \citet{2025arXiv251110582S_LYRA}. On the other hand, 1--2\,\% of the viral radius at $z=0$, $R_{\rm vir}(z=0)$, was often used in previous studies \citep{Cintio_2014,Tollet_2016,Lazar_2020}. Figure~\ref{fig:DMRadialDensity2} presents the power-law index of our DM halos fitted at 1--2\,\% of $R_{\rm vir}(z=0)$, similar to the previous studies. The $x$ range of the lines indicates $R_{\rm vir}(z=0)$.  
As seen in this figure, 1--2\,\% of $R_{\rm vir}(z=0)$ is outside the central cusp region, and therefore the fitted slope is much steeper than the values we obtained for 1--2\,$R_{\rm conv}$, except for the slope of the Halo 284 Hydro model, which is slightly steeper at 1--2\,$R_{\rm conv}$ than that at 1--2\,\% of $R_{\rm vir}(z=0)$.
For Halo 230, 1--2\,\% of $R_{\rm vir}(z=0)$ is obviously outside the central cusp. 
From these results, we adopted 1--2\,$R_{\rm conv}$ to measure the power-law slope of the cusp. 
The resolution of \citet{2025arXiv251110582S_LYRA} was higher than ours. In such cases, 1--2\,\% of $R_{\rm vir}$ is far from the convergence radius. Therefore, 1--2\,\% of $R_{\rm vir}$ is not sufficiently small to capture the power of the central cusp. However, the power measured at 1--2\,$R_{\rm conv}$ changes as the convergence radius changes with time, as shown in Fig.~\ref{fig:SlopeHistory_z0}.

\section{Previous simulations}
\label{sec:previous_sim}

Figure~\ref{fig:MstarMhalo_Alpha} plots the relation between the stellar-to-halo mass ratio and the power-law slope of the DM halo obtained from previous numerical studies \citep{2014MNRAS.437..415D,Tollet_2016,Lazar_2020} as the green, blue, and red shaded regions, respectively. These regions were obtained by fitting their results using the following functions:
\begin{equation}
    \alpha(x) = n-\log_{10} \left[\left(\frac{x}{x_0}\right)^{-\beta} + \left(\frac{x}{x_0}\right)^{\gamma}\right]
    \label{equation:MstarMhalo_alpha_DC14}
\end{equation}
and
\begin{equation}
    \alpha(x) = n-\log_{10} \left[n_1 \left(1+\frac{x}{x_1}\right)^{-\beta} + \left(\frac{x}{x_0}\right)^{\gamma}\right]
    \label{equation:MstarMhalo_alpha_other}
\end{equation}
where $x = M_{\star}/M_{\mathrm{halo}}$, and $n, n_1, x_0, x_1, \beta, \gamma$ are fitting parameters.
We used eq. (\ref{equation:MstarMhalo_alpha_DC14}) for the result of \citet{Cintio_2014} and eq. (\ref{equation:MstarMhalo_alpha_other}) for those of \citet{Tollet_2016} and \citet{Lazar_2020}. 
The fitting parameters are summarized in Table \ref{table:MstarMhalo_alpha}, and we have taken $\Delta \alpha = 0.2$ for all vertical widths.

\begin{table*}
 \caption{Parameter values of the fitting functions for previous simulations.}
 \label{table:MstarMhalo_alpha}
 \centering
  \begin{tabular}{cccccccc}
   \hline
   Paper & Function & $n$ & $n_1$ & $x_0$ & $x_1$ & $\beta$ & $\gamma$ \\
   \hline \hline
   \citet{Cintio_2014} & eq. \ref{equation:MstarMhalo_alpha_DC14} & $0.132$ & - & $8.89\times10^{-3}$ & - & $0.593$ & $1.99$ \\
   \citet{Tollet_2016} & eq. \ref{equation:MstarMhalo_alpha_other}& $-0.158$ & $26.49$ & $8.77\times10^{-3}$ & $9.44\times10^{-5}$ & $0.85$ & $1.66$ \\
   \citet{Lazar_2020} & eq. \ref{equation:MstarMhalo_alpha_other}& $-1.60$ & $0.80$ & $9.18\times10^{-2}$ & $6.54\times10^{-3}$ & $5$ & $1.05$ \\
   \hline
  \end{tabular}
\end{table*}

\bibliographystyle{aasjournal}
\bibliography{bibs}

@ARTICLE{2015MNRAS.452.3650O,
       author = {{Oman}, Kyle A. and {Navarro}, Julio F. and {Fattahi}, Azadeh and {Frenk}, Carlos S. and {Sawala}, Till and {White}, Simon D.~M. and {Bower}, Richard and {Crain}, Robert A. and {Furlong}, Michelle and {Schaller}, Matthieu and {Schaye}, Joop and {Theuns}, Tom},
        title = "{The unexpected diversity of dwarf galaxy rotation curves}",
      journal = {\mnras},
         year = 2015,
        month = oct,
       volume = {452},
       number = {4},
        pages = {3650-3665},
          doi = {10.1093/mnras/stv1504},
archivePrefix = {arXiv},
       eprint = {1504.01437},
 primaryClass = {astro-ph.GA},
       adsurl = {https://ui.adsabs.harvard.edu/abs/2015MNRAS.452.3650O}
}

@ARTICLE{2021MNRAS.504.3509O_EDGE,
       author = {{Orkney}, Matthew D.~A. and {Read}, Justin I. and {Rey}, Martin P. and {Nasim}, Imran and {Pontzen}, Andrew and {Agertz}, Oscar and {Kim}, Stacy Y. and {Delorme}, Maxime and {Dehnen}, Walter},
        title = "{EDGE: two routes to dark matter core formation in ultra-faint dwarfs}",
      journal = {\mnras},
         year = 2021,
        month = jul,
       volume = {504},
       number = {3},
        pages = {3509-3522},
          doi = {10.1093/mnras/stab1066},
archivePrefix = {arXiv},
       eprint = {2101.02688},
 primaryClass = {astro-ph.GA},
       adsurl = {https://ui.adsabs.harvard.edu/abs/2021MNRAS.504.3509O}
}

@ARTICLE{2025arXiv251110582S_LYRA,
       author = {{Sureda}, Joaquin and {Brown}, Shaun T. and {Fattahi}, Azadeh and {Gutcke}, Thales and {Bose}, Sownak and {Doppel}, Jessica E. and {Pakmor}, R{\"u}diger},
        title = "{Co-evolution of baryons and dark matter halos of LYRA dwarf galaxies}",
      journal = {arXiv e-prints},
         year = 2025,
        month = nov,
          eid = {arXiv:2511.10582},
        pages = {arXiv:2511.10582},
          doi = {10.48550/arXiv.2511.10582},
archivePrefix = {arXiv},
       eprint = {2511.10582},
 primaryClass = {astro-ph.GA},
       adsurl = {https://ui.adsabs.harvard.edu/abs/2025arXiv251110582S}
}

@ARTICLE{GEAR,
       author = {{Revaz}, Yves and {Jablonka}, Pascale},
        title = "{Pushing back the limits: detailed properties of dwarf galaxies in a {\ensuremath{\Lambda}}CDM universe}",
      journal = {\aap},
         year = 2018,
        month = aug,
       volume = {616},
          eid = {A96},
        pages = {A96},
          doi = {10.1051/0004-6361/201832669},
archivePrefix = {arXiv},
       eprint = {1801.06222},
 primaryClass = {astro-ph.GA},
       adsurl = {https://ui.adsabs.harvard.edu/abs/2018A&A...616A..96R}
}

@ARTICLE{LYRA,
       author = {{Gutcke}, Thales A. and {Pakmor}, R{\"u}diger and {Naab}, Thorsten and {Springel}, Volker},
        title = "{LYRA - I. Simulating the multiphase ISM of a dwarf galaxy with variable energy supernovae from individual stars}",
      journal = {\mnras},
         year = 2021,
        month = mar,
       volume = {501},
       number = {4},
        pages = {5597-5615},
          doi = {10.1093/mnras/staa3875},
archivePrefix = {arXiv},
       eprint = {2010.07311},
 primaryClass = {astro-ph.GA},
       adsurl = {https://ui.adsabs.harvard.edu/abs/2021MNRAS.501.5597G}
}

@ARTICLE{2022NatAs...6..897S,
       author = {{Sales}, Laura V. and {Wetzel}, Andrew and {Fattahi}, Azadeh},
        title = "{Baryonic solutions and challenges for cosmological models of dwarf galaxies}",
      journal = {Nature Astronomy},
         year = 2022,
        month = jun,
       volume = {6},
        pages = {897-910},
          doi = {10.1038/s41550-022-01689-w},
archivePrefix = {arXiv},
       eprint = {2206.05295},
 primaryClass = {astro-ph.GA},
       adsurl = {https://ui.adsabs.harvard.edu/abs/2022NatAs...6..897S}
}

@ARTICLE{2013MNRAS.428.3121M,
       author = {{Moster}, Benjamin P. and {Naab}, Thorsten and {White}, Simon D.~M.},
        title = "{Galactic star formation and accretion histories from matching galaxies to dark matter haloes}",
      journal = {\mnras},
         year = 2013,
        month = feb,
       volume = {428},
       number = {4},
        pages = {3121-3138},
          doi = {10.1093/mnras/sts261},
archivePrefix = {arXiv},
       eprint = {1205.5807},
 primaryClass = {astro-ph.CO},
       adsurl = {https://ui.adsabs.harvard.edu/abs/2013MNRAS.428.3121M}
}

@ARTICLE{2013ApJ...770...57B,
       author = {{Behroozi}, Peter S. and {Wechsler}, Risa H. and {Conroy}, Charlie},
        title = "{The Average Star Formation Histories of Galaxies in Dark Matter Halos from z = 0-8}",
      journal = {\apj},
         year = 2013,
        month = jun,
       volume = {770},
       number = {1},
          eid = {57},
        pages = {57},
          doi = {10.1088/0004-637X/770/1/57},
archivePrefix = {arXiv},
       eprint = {1207.6105},
 primaryClass = {astro-ph.CO},
       adsurl = {https://ui.adsabs.harvard.edu/abs/2013ApJ...770...57B}
}

@ARTICLE{2019MNRAS.483.3363H,
       author = {{Hu}, Chia-Yu},
        title = "{Supernova-driven winds in simulated dwarf galaxies}",
      journal = {\mnras},
         year = 2019,
        month = mar,
       volume = {483},
       number = {3},
        pages = {3363-3381},
          doi = {10.1093/mnras/sty3252},
archivePrefix = {arXiv},
       eprint = {1805.06614},
 primaryClass = {astro-ph.GA},
       adsurl = {https://ui.adsabs.harvard.edu/abs/2019MNRAS.483.3363H}
}

@ARTICLE{2020MNRAS.491.1656A,
       author = {{Agertz}, Oscar and {Pontzen}, Andrew and {Read}, Justin I. and {Rey}, Martin P. and {Orkney}, Matthew and {Rosdahl}, Joakim and {Teyssier}, Romain and {Verbeke}, Robbert and {Kretschmer}, Michael and {Nickerson}, Sarah},
        title = "{EDGE: the mass-metallicity relation as a critical test of galaxy formation physics}",
      journal = {\mnras},
         year = 2020,
        month = jan,
       volume = {491},
       number = {2},
        pages = {1656-1672},
          doi = {10.1093/mnras/stz3053},
archivePrefix = {arXiv},
       eprint = {1904.02723},
 primaryClass = {astro-ph.GA},
       adsurl = {https://ui.adsabs.harvard.edu/abs/2020MNRAS.491.1656A}
}

@ARTICLE{2025MNRAS.536..314M,
       author = {{Muni}, Claudia and {Pontzen}, Andrew and {Read}, Justin I. and {Agertz}, Oscar and {Rey}, Martin P. and {Taylor}, Ethan and {Kim}, Stacy Y. and {Gray}, Emily I.},
        title = "{EDGE: dark matter core creation depends on the timing of star formation}",
      journal = {\mnras},
         year = 2025,
        month = jan,
       volume = {536},
       number = {1},
        pages = {314-323},
          doi = {10.1093/mnras/stae2748},
archivePrefix = {arXiv},
       eprint = {2407.14579},
 primaryClass = {astro-ph.GA},
       adsurl = {https://ui.adsabs.harvard.edu/abs/2025MNRAS.536..314M}
}

@ARTICLE{2025arXiv250722155H,
       author = {{Hayashi}, Kohei and {Kaneda}, Yuka and {Mori}, Masao and {Shinozaki}, Michi},
        title = "{Cusp-to-Core Transition of Dark Matter Halos across Galaxy Mass Scales}",
      journal = {arXiv e-prints},
         year = 2025,
        month = jul,
          eid = {arXiv:2507.22155},
        pages = {arXiv:2507.22155},
          doi = {10.48550/arXiv.2507.22155},
archivePrefix = {arXiv},
       eprint = {2507.22155},
 primaryClass = {astro-ph.GA},
       adsurl = {https://ui.adsabs.harvard.edu/abs/2025arXiv250722155H}
}

@ARTICLE{2023ApJ...953..185H,
       author = {{Hayashi}, Kohei and {Hirai}, Yutaka and {Chiba}, Masashi and {Ishiyama}, Tomoaki},
        title = "{Dark Matter Halo Properties of the Galactic Dwarf Satellites: Implication for Chemo-dynamical Evolution of the Satellites and a Challenge to Lambda Cold Dark Matter}",
      journal = {\apj},
         year = 2023,
        month = aug,
       volume = {953},
       number = {2},
          eid = {185},
        pages = {185},
          doi = {10.3847/1538-4357/ace33e},
archivePrefix = {arXiv},
       eprint = {2206.02821},
 primaryClass = {astro-ph.GA},
       adsurl = {https://ui.adsabs.harvard.edu/abs/2023ApJ...953..185H}
}

@ARTICLE{2024PASJ...76.1026K,
       author = {{Kaneda}, Yuka and {Mori}, Masao and {Otaki}, Koki},
        title = "{A universal scaling relation incorporating the cusp-to-core transition of dark matter halos}",
      journal = {\pasj},
         year = 2024,
        month = oct,
       volume = {76},
       number = {5},
        pages = {1026-1040},
          doi = {10.1093/pasj/psae068},
archivePrefix = {arXiv},
       eprint = {2407.03614},
 primaryClass = {astro-ph.GA},
       adsurl = {https://ui.adsabs.harvard.edu/abs/2024PASJ...76.1026K}
}

@ARTICLE{2020ApJS..250...17L,
       author = {{Liu}, Chengze and {C{\^o}t{\'e}}, Patrick and {Peng}, Eric W. and {Roediger}, Joel and {Zhang}, Hongxin and {Ferrarese}, Laura and {S{\'a}nchez-Janssen}, Ruben and {Guhathakurta}, Puragra and {Yang}, Xiaohu and {Jing}, Yipeng and {Alamo-Mart{\'\i}nez}, Karla and {Blakeslee}, John P. and {Boselli}, Alessandro and {Cuilandre}, Jean-Charles and {Duc}, Pierre-Alain and {Durrell}, Patrick and {Gwyn}, Stephen and {Jord{\'a}n}, Andres and {Ko}, Youkyung and {Lan{\c{c}}on}, Ariane and {Lim}, Sungsoon and {Longobardi}, Alessia and {Mei}, Simona and {Mihos}, J. Christopher and {Mu{\~n}oz}, Roberto and {Powalka}, Mathieu and {Puzia}, Thomas and {Spengler}, Chelsea and {Toloba}, Elisa},
        title = "{The Next Generation Virgo Cluster Survey. XXXIV. Ultracompact Dwarf Galaxies in the Virgo Cluster}",
      journal = {\apjs},
         year = 2020,
        month = sep,
       volume = {250},
       number = {1},
          eid = {17},
        pages = {17},
          doi = {10.3847/1538-4365/abad91},
archivePrefix = {arXiv},
       eprint = {2007.15275},
 primaryClass = {astro-ph.GA},
       adsurl = {https://ui.adsabs.harvard.edu/abs/2020ApJS..250...17L}
}

@ARTICLE{2018MNRAS.478.1520B,
       author = {{Baumgardt}, H. and {Hilker}, M.},
        title = "{A catalogue of masses, structural parameters, and velocity dispersion profiles of 112 Milky Way globular clusters}",
      journal = {\mnras},
         year = 2018,
        month = aug,
       volume = {478},
       number = {2},
        pages = {1520-1557},
          doi = {10.1093/mnras/sty1057},
archivePrefix = {arXiv},
       eprint = {1804.08359},
 primaryClass = {astro-ph.GA},
       adsurl = {https://ui.adsabs.harvard.edu/abs/2018MNRAS.478.1520B}
}

@ARTICLE{2012AJ....144....4M,
       author = {{McConnachie}, Alan W.},
        title = "{The Observed Properties of Dwarf Galaxies in and around the Local Group}",
      journal = {\aj},
         year = 2012,
        month = jul,
       volume = {144},
       number = {1},
          eid = {4},
        pages = {4},
          doi = {10.1088/0004-6256/144/1/4},
archivePrefix = {arXiv},
       eprint = {1204.1562},
 primaryClass = {astro-ph.CO},
       adsurl = {https://ui.adsabs.harvard.edu/abs/2012AJ....144....4M}
}

@ARTICLE{2014MNRAS.437..415D,
       author = {{Di Cintio}, Arianna and {Brook}, Chris B. and {Macci{\`o}}, Andrea V. and {Stinson}, Greg S. and {Knebe}, Alexander and {Dutton}, Aaron A. and {Wadsley}, James},
        title = "{The dependence of dark matter profiles on the stellar-to-halo mass ratio: a prediction for cusps versus cores}",
      journal = {\mnras},
         year = 2014,
        month = jan,
       volume = {437},
       number = {1},
        pages = {415-423},
          doi = {10.1093/mnras/stt1891},
archivePrefix = {arXiv},
       eprint = {1306.0898},
 primaryClass = {astro-ph.CO},
       adsurl = {https://ui.adsabs.harvard.edu/abs/2014MNRAS.437..415D}
}

@ARTICLE{2014ApJ...793...46O,
       author = {{Ogiya}, Go and {Mori}, Masao},
        title = "{The Core-Cusp Problem in Cold Dark Matter Halos and Supernova Feedback: Effects of Oscillation}",
      journal = {\apj},
         year = 2014,
        month = sep,
       volume = {793},
       number = {1},
          eid = {46},
        pages = {46},
          doi = {10.1088/0004-637X/793/1/46},
archivePrefix = {arXiv},
       eprint = {1206.5412},
 primaryClass = {astro-ph.CO},
       adsurl = {https://ui.adsabs.harvard.edu/abs/2014ApJ...793...46O}
}

@ARTICLE{2017ARA&A..55..343B,
       author = {{Bullock}, James S. and {Boylan-Kolchin}, Michael},
        title = "{Small-Scale Challenges to the {\ensuremath{\Lambda}}CDM Paradigm}",
      journal = {\araa},
         year = 2017,
        month = aug,
       volume = {55},
       number = {1},
        pages = {343-387},
          doi = {10.1146/annurev-astro-091916-055313},
archivePrefix = {arXiv},
       eprint = {1707.04256},
 primaryClass = {astro-ph.CO},
       adsurl = {https://ui.adsabs.harvard.edu/abs/2017ARA&A..55..343B}
}

@ARTICLE{2009MNRAS.399L.174O,
       author = {{Okamoto}, Takashi and {Frenk}, Carlos S.},
        title = "{The origin of failed subhaloes and the common mass scale of the Milky Way satellite galaxies}",
      journal = {\mnras},
         year = 2009,
        month = oct,
       volume = {399},
       number = {1},
        pages = {L174-L178},
          doi = {10.1111/j.1745-3933.2009.00748.x},
archivePrefix = {arXiv},
       eprint = {0909.0262},
 primaryClass = {astro-ph.CO},
       adsurl = {https://ui.adsabs.harvard.edu/abs/2009MNRAS.399L.174O}
}

@ARTICLE{2012MNRAS.421.3464P,
       author = {{Pontzen}, Andrew and {Governato}, Fabio},
        title = "{How supernova feedback turns dark matter cusps into cores}",
      journal = {\mnras},
         year = 2012,
        month = apr,
       volume = {421},
       number = {4},
        pages = {3464-3471},
          doi = {10.1111/j.1365-2966.2012.20571.x},
archivePrefix = {arXiv},
       eprint = {1106.0499},
 primaryClass = {astro-ph.CO},
       adsurl = {https://ui.adsabs.harvard.edu/abs/2012MNRAS.421.3464P}
}

@ARTICLE{1995ApJ...447L..25B,
       author = {{Burkert}, A.},
        title = "{The Structure of Dark Matter Halos in Dwarf Galaxies}",
      journal = {\apjl},
         year = 1995,
        month = jul,
       volume = {447},
        pages = {L25-L28},
          doi = {10.1086/309560},
archivePrefix = {arXiv},
       eprint = {astro-ph/9504041},
 primaryClass = {astro-ph},
       adsurl = {https://ui.adsabs.harvard.edu/abs/1995ApJ...447L..25B}
}

@article{Tegmark_2004,
	doi = {10.1086/382125},
	url = {https://doi.org/10.1086/382125},
	year = 2004,
	month = {may},
	publisher = {American Astronomical Society},
	volume = {606},
	number = {2},
	pages = {702--740},
	author = {Max Tegmark and Michael R. Blanton and Michael A. Strauss and Fiona Hoyle and David Schlegel and Roman Scoccimarro and Michael S. Vogeley and David H. Weinberg and Idit Zehavi and Andreas Berlind and Tamas Budavari and Andrew Connolly and Daniel J. Eisenstein and Douglas Finkbeiner and Joshua A. Frieman and James E. Gunn and Andrew J. S. Hamilton and Lam Hui and Bhuvnesh Jain and David Johnston and Stephen Kent and Huan Lin and Reiko Nakajima and Robert C. Nichol and Jeremiah P. Ostriker and Adrian Pope and Ryan Scranton and Uro{\v{s}} Seljak and Ravi K. Sheth and Albert Stebbins and Alexander S. Szalay and Istvan Szapudi and Licia Verde and Yongzhong Xu and James Annis and Neta A. Bahcall and J. Brinkmann and Scott Burles and Francisco J. Castander and Istvan Csabai and Jon Loveday and Mamoru Doi and Masataka Fukugita and J. Richard Gott III and Greg Hennessy and David W. Hogg and {\v{Z}}eljko Ivezi{\'{c}} and Gillian R. Knapp and Don Q. Lamb and Brian C. Lee and Robert H. Lupton and Timothy A. McKay and Peter Kunszt and Jeffrey A. Munn and Liam O'Connell and John Peoples and Jeffrey R. Pier and Michael Richmond and Constance Rockosi and Donald P. Schneider and Christopher Stoughton and Douglas L. Tucker and Daniel E. Vanden Berk and Brian Yanny and Donald G. York and},
	title = {The Three-Dimensional Power Spectrum of Galaxies from the Sloan Digital Sky Survey},
	journal = {The Astrophysical Journal}
}

@article{Navarro_1996,
	doi = {10.1086/177173},
	url = {https://doi.org/10.1086/177173},
	year = 1996,
	month = {may},
	publisher = {American Astronomical Society},
	volume = {462},
	pages = {563},
	author = {Julio F. Navarro and Carlos S. Frenk and Simon D. M. White},
	title = {The Structure of Cold Dark Matter Halos},
	journal = {The Astrophysical Journal}
}

@article{Wheeler_2019,
	doi = {10.1093/mnras/stz2887},
	url = {https://doi.org/10.1093/mnras/stz2887},
	year = 2019,
	month = {oct},
	publisher = {Oxford University Press ({OUP})},
	volume = {490},
	number = {3},
	pages = {4447--4463},
	author = {Coral Wheeler and Philip F Hopkins and Andrew B Pace and Shea Garrison-Kimmel and Michael Boylan-Kolchin and Andrew Wetzel and James S Bullock and Du{\v{s}}an Kere{\v{s}} and Claude-Andr{\'{e}} Faucher-Gigu{\`{e}}re and Eliot Quataert},
	title = {Be it therefore resolved: cosmological simulations of dwarf galaxies with 30 solar mass resolution},
	journal = {Monthly Notices of the Royal Astronomical Society}
}

@article{Power_2003,
	doi = {10.1046/j.1365-8711.2003.05925.x},
	url = {https://doi.org/10.1046/j.1365-8711.2003.05925.x},
	year = 2003,
	month = {jan},
	publisher = {Oxford University Press ({OUP})},
	volume = {338},
	number = {1},
	pages = {14--34},
	author = {C. Power and J. F. Navarro and A. Jenkins and C. S. Frenk and S. D. M. White and V. Springel and J. Stadel and T. Quinn},
	title = {The inner structure of $\mathrm{\Lambda}$CDM haloes -- I. A numerical convergence study},
	journal = {Monthly Notices of the Royal Astronomical Society}
}

@article{Saitoh_2009,
	doi = {10.1093/pasj/61.3.481},
	url = {https://doi.org/10.1093/pasj/61.3.481},
	year = 2009,
	month = {jun},
	publisher = {Oxford University Press ({OUP})},
	volume = {61},
	number = {3},
	pages = {481--486},
	author = {Takayuki R. Saitoh and Hiroshi Daisaka and Eiichiro Kokubo and Junichiro Makino and Takashi Okamoto and Kohji Tomisaka and Keiichi Wada and Naoki Yoshida},
	title = {Toward First-Principle Simulations of Galaxy Formation: {II}. Shock-Induced Starburst at a Collision Interface during the First Encounter of Interacting Galaxies},
	journal = {Publications of the Astronomical Society of Japan}
}

@article{Hirai_2021,
	doi = {10.1093/pasj/psab038},
	url = {https://doi.org/10.1093/pasj/psab038},
	year = 2021,
	month = {may},
	publisher = {Oxford University Press ({OUP})},
	volume = {73},
	number = {4},
	pages = {1036--1056},
	author = {Yutaka Hirai and Michiko S Fujii and Takayuki R Saitoh},
	title = {{SIRIUS} project. I. Star formation models for star-by-star simulations of star clusters and galaxy formation},
	journal = {Publications of the Astronomical Society of Japan}
}

@ARTICLE{Fujii_2021,
       author = {{Fujii}, Michiko S. and {Saitoh}, Takayuki R. and {Hirai}, Yutaka and {Wang}, Long},
        title = "{SIRIUS project. III. Star-by-star simulations of star cluster formation using a direct N-body integrator with stellar feedback}",
      journal = {\pasj},
         year = 2021,
        month = aug,
       volume = {73},
       number = {4},
        pages = {1074-1099},
          doi = {10.1093/pasj/psab061},
archivePrefix = {arXiv},
       eprint = {2103.02829},
 primaryClass = {astro-ph.GA},
       adsurl = {https://ui.adsabs.harvard.edu/abs/2021PASJ...73.1074F}
}

@ARTICLE{Hirai_2025,
       author = {{Hirai}, Yutaka and {Saitoh}, Takayuki R. and {Fujii}, Michiko S. and {Kaneko}, Katsuhiro and {Beers}, Timothy C.},
        title = "{SIRIUS: Identifying Metal-poor Stars Enriched by a Single Supernova in a Dwarf Galaxy Cosmological Zoom-in Simulation Resolving Individual Massive Stars}",
      journal = {\apjl},
         year = 2025,
        month = feb,
       volume = {980},
       number = {2},
          eid = {L25},
        pages = {L25},
          doi = {10.3847/2041-8213/adaf95},
archivePrefix = {arXiv},
       eprint = {2411.18680},
 primaryClass = {astro-ph.GA},
       adsurl = {https://ui.adsabs.harvard.edu/abs/2025ApJ...980L..25H}
}

@article{Saitoh_2017,
	doi = {10.3847/1538-3881/153/2/85},
	url = {https://doi.org/10.3847/1538-3881/153/2/85},
	year = 2017,
	month = {jan},
	publisher = {American Astronomical Society},
	volume = {153},
	number = {2},
	pages = {85},
	author = {Takayuki R. Saitoh},
	title = {{CHEMICAL} {EVOLUTION} {LIBRARY} {FOR} {GALAXY} {FORMATION} {SIMULATION}},
	journal = {The Astronomical Journal}
}

@article{Hopkins_2018,
	doi = {10.1093/mnras/sty1690},
	url = {https://doi.org/10.1093/mnras/sty1690},
	year = 2018,
	month = {jun},
	publisher = {Oxford University Press ({OUP})},
	volume = {480},
	number = {1},
	pages = {800--863},
	author = {Philip F Hopkins and Andrew Wetzel and Du{\v{s}}an Kere{\v{s}} and Claude-Andr{\'{e}} Faucher-Gigu{\`{e}}re and Eliot Quataert and Michael Boylan-Kolchin and Norman Murray and Christopher C Hayward and Shea Garrison-Kimmel and Cameron Hummels and Robert Feldmann and Paul Torrey and Xiangcheng Ma and Daniel Angl{\'{e}}s-Alc{\'{a}}zar and Kung-Yi Su and Matthew Orr and Denise Schmitz and Ivanna Escala and Robyn Sanderson and Michael Y Grudi{\'{c}} and Zachary Hafen and Ji-Hoon Kim and Alex Fitts and James S Bullock and Coral Wheeler and T K Chan and Oliver D Elbert and Desika Narayanan},
	title = {{FIRE}-2 simulations: physics versus numerics in galaxy formation},
	journal = {Monthly Notices of the Royal Astronomical Society}
}

@article{Planck_2020,
	doi = {10.1051/0004-6361/201833910},
	url = {https://doi.org/10.1051/0004-6361/201833910},
	year = 2020,
	month = {sep},
	publisher = {{EDP} Sciences},
	volume = {641},
	pages = {A6},
	author = {{Planck Collaboration} and N. Aghanim and Y. Akrami and M. Ashdown and J. Aumont and C. Baccigalupi and M. Ballardini and A. J. Banday and R. B. Barreiro and N. Bartolo and S. Basak and R. Battye and K. Benabed and J.-P. Bernard and M. Bersanelli and P. Bielewicz and J. J. Bock and J. R. Bond and J. Borrill and F. R. Bouchet and F. Boulanger and M. Bucher and C. Burigana and R. C. Butler and E. Calabrese and J.-F. Cardoso and J. Carron and A. Challinor and H. C. Chiang and J. Chluba and L. P. L. Colombo and C. Combet and D. Contreras and B. P. Crill and F. Cuttaia and P. de Bernardis and G. de Zotti and J. Delabrouille and J.-M. Delouis and E. Di Valentino and J. M. Diego and O. Dor{\'{e}} and M. Douspis and A. Ducout and X. Dupac and S. Dusini and G. Efstathiou and F. Elsner and T. A. En{\ss}lin and H. K. Eriksen and Y. Fantaye and M. Farhang and J. Fergusson and R. Fernandez-Cobos and F. Finelli and F. Forastieri and M. Frailis and A. A. Fraisse and E. Franceschi and A. Frolov and S. Galeotta and S. Galli and K. Ganga and R. T. G{\'{e}}nova-Santos and M. Gerbino and T. Ghosh and J. Gonz{\'{a}}lez-Nuevo and K. M. G{\'{o}}rski and S. Gratton and A. Gruppuso and J. E. Gudmundsson and J. Hamann and W. Handley and F. K. Hansen and D. Herranz and S. R. Hildebrandt and E. Hivon and Z. Huang and A. H. Jaffe and W. C. Jones and A. Karakci and E. Keihänen and R. Keskitalo and K. Kiiveri and J. Kim and T. S. Kisner and L. Knox and N. Krachmalnicoff and M. Kunz and H. Kurki-Suonio and G. Lagache and J.-M. Lamarre and A. Lasenby and M. Lattanzi and C. R. Lawrence and M. Le Jeune and P. Lemos and J. Lesgourgues and F. Levrier and A. Lewis and M. Liguori and P. B. Lilje and M. Lilley and V. Lindholm and M. L{\'{o}}pez-Caniego and P. M. Lubin and Y.-Z. Ma and J. F. Mac{\'{\i}}as-P{\'{e}}rez and G. Maggio and D. Maino and N. Mandolesi and A. Mangilli and A. Marcos-Caballero and M. Maris and P. G. Martin and M. Martinelli and E. Mart{\'{\i}}nez-Gonz{\'{a}}lez and S. Matarrese and N. Mauri and J. D. McEwen and P. R. Meinhold and A. Melchiorri and A. Mennella and M. Migliaccio and M. Millea and S. Mitra and M.-A. Miville-Desch{\^{e}}nes and D. Molinari and L. Montier and G. Morgante and A. Moss and P. Natoli and H. U. N{\o}rgaard-Nielsen and L. Pagano and D. Paoletti and B. Partridge and G. Patanchon and H. V. Peiris and F. Perrotta and V. Pettorino and F. Piacentini and L. Polastri and G. Polenta and J.-L. Puget and J. P. Rachen and M. Reinecke and M. Remazeilles and A. Renzi and G. Rocha and C. Rosset and G. Roudier and J. A. Rubi{\~{n}}o-Mart{\'{\i}}n and B. Ruiz-Granados and L. Salvati and M. Sandri and M. Savelainen and D. Scott and E. P. S. Shellard and C. Sirignano and G. Sirri and L. D. Spencer and R. Sunyaev and A.-S. Suur-Uski and J. A. Tauber and D. Tavagnacco and M. Tenti and L. Toffolatti and M. Tomasi and T. Trombetti and L. Valenziano and J. Valiviita and B. Van Tent and L. Vibert and P. Vielva and F. Villa and N. Vittorio and B. D. Wandelt and I. K. Wehus and M. White and S. D. M. White and A. Zacchei and A. Zonca},
	title = {Planck 2018 results - VI. Cosmological parameters},
	journal = {Astronomy \& Astrophysics}
}

@article{Hahn_2011,
	doi = {10.1111/j.1365-2966.2011.18820.x},
	url = {https://doi.org/10.1111/j.1365-2966.2011.18820.x},
	year = 2011,
	month = {jul},
	publisher = {Oxford University Press ({OUP})},
	volume = {415},
	number = {3},
	pages = {2101--2121},
	author = {Oliver Hahn and Tom Abel},
	title = {Multi-scale initial conditions for cosmological simulations},
	journal = {Monthly Notices of the Royal Astronomical Society}
}

@article{Springel_2005b,
	doi = {10.1111/j.1365-2966.2005.09655.x},
	url = {https://doi.org/10.1111/j.1365-2966.2005.09655.x},
	year = 2005,
	month = {dec},
	publisher = {Oxford University Press ({OUP})},
	volume = {364},
	number = {4},
	pages = {1105--1134},
	author = {Volker Springel},
	title = {The cosmological simulation code gadget-2},
	journal = {Monthly Notices of the Royal Astronomical Society}
}

@article{Knollmann_2009,
	doi = {10.1088/0067-0049/182/2/608},
	url = {https://doi.org/10.1088/0067-0049/182/2/608},
	year = 2009,
	month = {may},
	publisher = {American Astronomical Society},
	volume = {182},
	number = {2},
	pages = {608--624},
	author = {Steffen R. Knollmann and Alexander Knebe},
	title = {{AHF}: {AMIGA}{\textquotesingle}S {HALO} {FINDER}},
	journal = {The Astrophysical Journal Supplement Series}
}

@article{Salpeter_1955,
	doi = {10.1086/145971},
	url = {https://doi.org/10.1086/145971},
	year = 1955,
	month = {jan},
	publisher = {American Astronomical Society},
	volume = {121},
	pages = {161},
	author = {Edwin E. Salpeter},
	title = {The Luminosity Function and Stellar Evolution.},
	journal = {The Astrophysical Journal}
}

@article{Springel_2005a,
  doi = {10.1038/nature03597},
  url = {https://doi.org/10.1038/nature03597},
  year = {2005},
  month = jun,
  publisher = {Springer Science and Business Media {LLC}},
  volume = {435},
  number = {7042},
  pages = {629--636},
  author = {Volker Springel and Simon D. M. White and Adrian Jenkins and Carlos S. Frenk and Naoki Yoshida and Liang Gao and Julio Navarro and Robert Thacker and Darren Croton and John Helly and John A. Peacock and Shaun Cole and Peter Thomas and Hugh Couchman and August Evrard and J\"{o}rg Colberg and Frazer Pearce},
  title = {Simulations of the formation,  evolution and clustering of galaxies and quasars},
  journal = {Nature}
}

@article{deBlok_2010,
  doi = {10.1155/2010/789293},
  url = {https://doi.org/10.1155/2010/789293},
  year = {2010},
  publisher = {Hindawi Limited},
  volume = {2010},
  pages = {1--14},
  author = {W. J. G. de Blok},
  title = {The Core-Cusp Problem},
  journal = {Advances in Astronomy}
}

@article{Flores_1994,
  doi = {10.1086/187350},
  url = {https://doi.org/10.1086/187350},
  year = {1994},
  month = may,
  publisher = {American Astronomical Society},
  volume = {427},
  pages = {L1},
  author = {Ricardo A. Flores and Joel R. Primack},
  title = {Observational and theoretical constraints on singular dark matter halos},
  journal = {The Astrophysical Journal}
}

@article{Moore_1994,
  doi = {10.1038/370629a0},
  url = {https://doi.org/10.1038/370629a0},
  year = {1994},
  month = aug,
  publisher = {Springer Science and Business Media {LLC}},
  volume = {370},
  number = {6491},
  pages = {629--631},
  author = {Ben Moore},
  title = {Evidence against dissipation-less dark matter from observations of galaxy haloes},
  journal = {Nature}
}

@article{Governato_2010,
  doi = {10.1038/nature08640},
  url = {https://doi.org/10.1038/nature08640},
  year = {2010},
  month = jan,
  publisher = {Springer Science and Business Media {LLC}},
  volume = {463},
  number = {7278},
  pages = {203--206},
  author = {F. Governato and C. Brook and L. Mayer and A. Brooks and G. Rhee and J. Wadsley and P. Jonsson and B. Willman and G. Stinson and T. Quinn and P. Madau},
  title = {Bulgeless dwarf galaxies and dark matter cores from supernova-driven outflows},
  journal = {Nature}
}

@article{Tollet_2016,
  doi = {10.1093/mnras/stv2856},
  url = {https://doi.org/10.1093/mnras/stv2856},
  year = {2016},
  month = jan,
  publisher = {Oxford University Press ({OUP})},
  volume = {456},
  number = {4},
  pages = {3542--3552},
  author = {Edouard Tollet and Andrea V. Macci{\`{o}} and Aaron A. Dutton and Greg S. Stinson and Liang Wang and Camilla Penzo and Thales A. Gutcke and Tobias Buck and Xi Kang and Chris Brook and Arianna Di Cintio and Ben W. Keller and James Wadsley},
  title = {{NIHAO} {\textendash} {IV}: core creation and destruction in dark matter density profiles across cosmic time},
  journal = {Monthly Notices of the Royal Astronomical Society}
}

@article{Cintio_2014,
  doi = {10.1093/mnras/stu729},
  url = {https://doi.org/10.1093/mnras/stu729},
  year = {2014},
  month = may,
  publisher = {Oxford University Press ({OUP})},
  volume = {441},
  number = {4},
  pages = {2986--2995},
  author = {Di Cintio, Arianna and Chris B. Brook and Aaron A. Dutton and Andrea V. Macci{\`{o}} and Greg S. Stinson and Alexander Knebe},
  title = {A mass-dependent density profile for dark matter haloes including the influence of galaxy formation},
  journal = {Monthly Notices of the Royal Astronomical Society}
}

@article{Hayashi_2020,
	doi = {10.3847/1538-4357/abbe0a},
	url = {https://doi.org/10.3847/1538-4357/abbe0a},
	year = 2020,
	month = {nov},
	publisher = {American Astronomical Society},
	volume = {904},
	number = {1},
	pages = {45},
	author = {Kohei Hayashi and Masashi Chiba and Tomoaki Ishiyama},
	title = {Diversity of Dark Matter Density Profiles in the Galactic Dwarf Spheroidal Satellites},
	journal = {The Astrophysical Journal}
}

@ARTICLE{2011AJ....141..193O,
       author = {{Oh}, Se-Heon and {de Blok}, W.~J.~G. and {Brinks}, Elias and {Walter}, Fabian and {Kennicutt}, Jr., Robert C.},
        title = "{Dark and Luminous Matter in THINGS Dwarf Galaxies}",
      journal = {\aj},
         year = 2011,
        month = jun,
       volume = {141},
       number = {6},
          eid = {193},
        pages = {193},
          doi = {10.1088/0004-6256/141/6/193},
archivePrefix = {arXiv},
       eprint = {1011.0899},
 primaryClass = {astro-ph.CO},
       adsurl = {https://ui.adsabs.harvard.edu/abs/2011AJ....141..193O}
}

@article{Fitts_2017,
	doi = {10.1093/mnras/stx1757},
	url = {https://doi.org/10.1093/mnras/stx1757},
	year = 2017,
	month = {jul},
	publisher = {Oxford University Press ({OUP})},
	volume = {471},
	number = {3},
	pages = {3547--3562},
	author = {Alex Fitts and Michael Boylan-Kolchin and Oliver D. Elbert and James S. Bullock and Philip F. Hopkins and Jose O{\~{n}}orbe and Andrew Wetzel and Coral Wheeler and Claude-Andr{\'{e}} Faucher-Gigu{\`{e}}re and Du{\v{s}}an Kere{\v{s}} and Evan D. Skillman and Daniel R. Weisz},
	title = {fire in the field: simulating the threshold of galaxy formation},
	journal = {Monthly Notices of the Royal Astronomical Society}
}

@article{Klypin_1997,
	doi = {10.48550/arXiv.astro-ph/9712217},
	url = {https://doi.org/10.48550/arXiv.astro-ph/9712217},
	year = 1997,
	month = {dec},
	publisher = {arXiv},
	author = {Klypin, Anatoly and Holtzman, Jon},
	title = {Particle-Mesh code for cosmological simulations},
	journal = {arXiv}
}

@article{Navarro_2004,
	doi = {10.1111/j.1365-2966.2004.07586.x},
	url = {https://doi.org/10.1111/j.1365-2966.2004.07586.x},
	year = 2004,
	month = {apr},
	publisher = {Oxford University Press ({OUP})},
	volume = {349},
	number = {3},
	pages = {1039--1051},
	author = {J. F. Navarro and E. Hayashi and C. Power and A. R. Jenkins and C. S. Frenk and S. D. M. White and V. Springel and J. Stadel and T. R. Quinn},
	title = {The inner structure of ΛCDM haloes - {III}. Universality and asymptotic slopes},
	journal = {Monthly Notices of the Royal Astronomical Society}
}

@article{Schaller_2015,
	doi = {10.1093/mnras/stv1067},
	url = {https://doi.org/10.1093/mnras/stv1067},
	year = 2015,
	month = {jun},
	publisher = {Oxford University Press ({OUP})},
	volume = {451},
	number = {2},
	pages = {1247--1267},
	author = {Matthieu Schaller and Carlos S. Frenk and Richard G. Bower and Tom Theuns and Adrian Jenkins and Joop Schaye and Robert A. Crain and Michelle Furlong and Claudio Dalla Vecchia and I. G. McCarthy},
	title = {Baryon effects on the internal structure of ΛCDM haloes in the {EAGLE} simulations},
	journal = {Monthly Notices of the Royal Astronomical Society}
}

@article{Lazar_2020,
	doi = {10.1093/mnras/staa2101},
	url = {https://doi.org/10.1093/mnras/staa2101},
	year = 2020,
	month = {jul},
	publisher = {Oxford University Press ({OUP})},
	volume = {497},
	number = {2},
	pages = {2393--2417},
	author = {Alexandres Lazar and James S Bullock and Michael Boylan-Kolchin and T K Chan and Philip F Hopkins and Andrew S Graus and Andrew Wetzel and Kareem El-Badry and Coral Wheeler and Maria C Straight and Du{\v{s}}an Kere{\v{s}} and Claude-Andr{\'{e}} Faucher-Gigu{\`{e}}re and Alex Fitts and Shea Garrison-Kimmel},
	title = {A dark matter profile to model diverse feedback-induced core sizes of ΛCDM haloes},
	journal = {Monthly Notices of the Royal Astronomical Society}
}

@article{Saitoh_2013,
	doi = {10.1088/0004-637x/768/1/44},
	url = {https://doi.org/10.1088/0004-637x/768%2F1%2F44},
	year = 2013,
	month = {apr},
	publisher = {American Astronomical Society},
	volume = {768},
	number = {1},
	pages = {44},
	author = {Takayuki R. Saitoh and Junichiro Makino},
	title = {A {DENSITY}-{INDEPENDENT} {FORMULATION} {OF} {SMOOTHED} {PARTICLE} {HYDRODYNAMICS}},
	journal = {The Astrophysical Journal}
}

@article{Barnes_1986,
	doi = {10.1038/324446a0},
	url = {https://doi.org/10.1038/324446a0},
	year = 1986,
	month = {dec},
	publisher = {Springer Science and Business Media {LLC}},
	volume = {324},
	number = {6096},
	pages = {446--449},
	author = {Josh Barnes and Piet Hut},
	title = {A hierarchical O(N log N) force-calculation algorithm},
	journal = {Nature}
}

@article{Ferland_1998,
	doi = {10.1086/316190},
	url = {https://doi.org/10.1086/316190},
	year = 1998,
	month = {jul},
	publisher = {{IOP} Publishing},
	volume = {110},
	number = {749},
	pages = {761--778},
	author = {G.~J.~ Ferland and K.~T.~ Korista and D.~A.~ Verner and J.~W.~ Ferguson and J.~B.~ Kingdon and E.~M.~ Verner},
	title = {{CLOUDY} 90: Numerical Simulation of Plasmas and Their Spectra},
	journal = {Publications of the Astronomical Society of the Pacific}
}

@article{Ferland_2013,
	doi = {arXiv:1302.4485},
	url = {https://doi.org/10.48550/arXiv.1302.4485},
	year = 2013,
	month = {apl},
	publisher = {Instituto de Astronom{\'\i}a},
	volume = {49},
	number={1},
	pages = {137--163},
	author = {G. J. Ferland and R. L. Porter and P. A. M. van Hoof and R. J. R. Williams and N. P. Abel and M. L. Lykins and Gargi Shaw and W. J. Henney and P. C. Stancil},
	title = {The 2013 Release of Cloudy},
	journal = {Revista mexicana de astronom{\'\i}a y astrof{\'\i}sica}}

@article{Ferland_2017,
	doi = {arXiv:1705.10877},
	url = {https://doi.org/10.48550/arXiv.1705.10877},
	year = 2017,
	month = {apl},
	publisher = {Instituto de Astronom{\'\i}a},
	volume = {53},
	number={2},
	pages = {385--438},
	author = {G. J. Ferland and M. Chatzikos and F. Guzm{\'a}n and M. L. Lykins and P. A. M. van Hoof and R. J. R. Williams and N. P. Abel and N. R. Badnell and F. P. Keenan and R. L. Porter and P. C. Stancil},
	title = {The 2017 Release of Cloudy},
	journal = {Revista mexicana de astronom{\'\i}a y astrof{\'\i}sica}
}

@article{Haardt_2012,
	doi = {10.1088/0004-637x/746/2/125},
	url = {https://doi.org/10.1088/0004-637x/746/2/125},
	year = 2012,
	month = {feb},
	publisher = {American Astronomical Society},
	volume = {746},
	number = {2},
	pages = {125},
	author = {Francesco Haardt and Piero Madau},
	title = {{RADIATIVE} {TRANSFER} {IN} A {CLUMPY} {UNIVERSE}. {IV}. {NEW} {SYNTHESIS} {MODELS} {OF} {THE} {COSMIC} {UV}/X-{RAY} {BACKGROUND}},
	journal = {The Astrophysical Journal}
}

@article{Rahmati_2013,
	doi = {10.1093/mnras/stt066},
	url = {https://doi.org/10.1093/mnras/stt066},
	year = 2013,
	month = {feb},
	publisher = {Oxford University Press ({OUP})},
	volume = {430},
	number = {3},
	pages = {2427--2445},
	author = {Alireza Rahmati and Andreas H. Pawlik and Milan Rai{\v{c}}ević and Joop Schaye},
	title = {On the evolution of the H{\hspace{0.167em}}i column density distribution in cosmological simulations},
	journal = {Monthly Notices of the Royal Astronomical Society}
}

@article{Schmidt_1959,
	doi = {10.1086/146614},
	url = {https://doi.org/10.1086/146614},
	year = 1959,
	month = {mar},
	publisher = {American Astronomical Society},
	volume = {129},
	pages = {243},
	author = {Maarten Schmidt},
	title = {The Rate of Star Formation.},
	journal = {The Astrophysical Journal}
}

@article{Katz_1992,
	doi = {10.1086/171366},
	url = {https://doi.org/10.1086/171366},
	year = 1992,
	month = {jun},
	publisher = {American Astronomical Society},
	volume = {391},
	pages = {502},
	author = {Neal Katz},
	title = {Dissipational galaxy formation. {II} - Effects of star formation},
	journal = {The Astrophysical Journal}
}

@article{Saitoh_2008,
	doi = {10.1093/pasj/60.4.667},
	url = {https://doi.org/10.1093/pasj/60.4.667},
	year = 2008,
	month = {aug},
	publisher = {Oxford University Press ({OUP})},
	volume = {60},
	number = {4},
	pages = {667--681},
	author = {Takayuki R. Saitoh and Hiroshi Daisaka and Eiichiro Kokubo and Junichiro Makino and Takashi Okamoto and Kohji Tomisaka and Keiichi Wada and Naoki Yoshida},
	title = {Toward First-Principle Simulations of Galaxy Formation: I. How Should We Choose Star-Formation Criteria in High-Resolution Simulations of Disk Galaxies?},
	journal = {Publications of the Astronomical Society of Japan}
}

@article{Portinari_1998,
    doi = {10.48550/arXiv.astro-ph/9711337},
    url = {https://doi.org/10.48550/arXiv.astro-ph/9711337},
    year = {1998},
    month = {jun},
    publisher = {{EDP} Sciences},
    volume = {334},
    pages = {505--539},
    author = {Portinari, L. and Chiosi, C. and Bressan, A.},
    title = {Galactic chemical enrichment with new metallicity dependent yields},
    journal = {Astronomy \& Astrophysics}
}

@article{Griffen_2016,
	doi = {10.3847/0004-637x/818/1/10},
	url = {https://doi.org/10.3847/0004-637x/818/1/10},
	year = 2016,
	month = {feb},
	publisher = {American Astronomical Society},
	volume = {818},
	number = {1},
	pages = {10},
	author = {Brendan F. Griffen and Alexander P. Ji and Gregory A. Dooley and Facundo A. G{\'{o}}mez and Mark Vogelsberger and Brian W. O'Shea and Anna Frebel},
	title = {{THE} {CATERPILLAR} {PROJECT}: A {LARGE} {SUITE} {OF} {MILKY} {WAY} {SIZED} {HALOS}},
	journal = {The Astrophysical Journal}
}

@article{Orkney_2021,
	doi = {10.1093/mnras/stab1066},
	url = {https://doi.org/10.1093/mnras/stab1066},
	year = 2021,
	month = {apr},
	publisher = {Oxford University Press ({OUP})},
	volume = {504},
	number = {3},
	pages = {3509--3522},
	author = {Matthew D A Orkney and Justin I Read and Martin P Rey and Imran Nasim and Andrew Pontzen and Oscar Agertz and Stacy Y Kim and Maxime Delorme and Walter Dehnen},
	title = {{EDGE}: two routes to dark matter core formation in ultra-faint dwarfs},
	journal = {Monthly Notices of the Royal Astronomical Society}
}

@article{Dutton_2020,
	doi = {10.1093/mnras/staa3028},
	url = {https://doi.org/10.1093/mnras/staa3028},
	year = 2020,
	month = {oct},
	publisher = {Oxford University Press ({OUP})},
	volume = {499},
	number = {2},
	pages = {2648--2661},
	author = {Aaron A Dutton and Tobias Buck and Andrea V Macci{\`{o}} and Keri L Dixon and Marvin Blank and Aura Obreja},
	title = {{NIHAO} {\textendash} {XXV}. Convergence in the cusp-core transformation of cold dark matter haloes at high star formation thresholds},
	journal = {Monthly Notices of the Royal Astronomical Society}
}

@ARTICLE{Nomoto_2013,
       author = {{Nomoto}, Ken'ichi and {Kobayashi}, Chiaki and {Tominaga}, Nozomu},
        title = "{Nucleosynthesis in Stars and the Chemical Enrichment of Galaxies}",
      journal = {\araa},
         year = 2013,
        month = aug,
       volume = {51},
       number = {1},
        pages = {457-509},
          doi = {10.1146/annurev-astro-082812-140956},
       adsurl = {https://ui.adsabs.harvard.edu/abs/2013ARA&A..51..457N}
}

\end{document}